%% file: main.tex
\documentclass{article}

\usepackage[pages=all, color=black, position={current page.south}, placement=bottom, scale=1, opacity=1, vshift=5mm]{background}

\usepackage[margin=1in]{geometry} % full-width

\usepackage{amsmath}

\usepackage{amsthm}
\usepackage{amsfonts}
\usepackage{array}
\usepackage{algorithm}
\usepackage{algorithmicx}
\usepackage{float}
\usepackage{subcaption}
\usepackage[pagewise]{lineno}

\usepackage[utf8]{inputenc}
\usepackage{hyperref}
\hypersetup{
	unicode,
	pdfauthor={Author One, Author Two, Author Three},
	pdftitle={Inverse Reinforcement Learning for Lane Change Prediction: A Comparative Study with Game-Theoretic Models},
	pdfsubject={Lane Change Prediction, Inverse Reinforcement Learning, Game Theory},
	pdfkeywords={IRL, lane change, game theory, autonomous vehicles, highD dataset},
	pdfproducer={LaTeX},
	pdfcreator={pdflatex}
}

\usepackage[sort&compress,numbers,square]{natbib}
\usepackage{array}
\usepackage{ragged2e}
\newcolumntype{L}[1]{>{\RaggedRight\arraybackslash}m{#1}}

\theoremstyle{plain}
\newtheorem{theorem}{Theorem}

\theoremstyle{definition}
\newtheorem{definition}[theorem]{Definition}

\usepackage{graphicx, color}
\graphicspath{{fig/}}
\usepackage{algorithm, algpseudocode} % use algorithm and algorithmicx for typesetting algorithms
\usepackage{mathrsfs} % for \mathscr command

\title{Game-Theoretic Inverse Reinforcement Learning for Modeling Competitive Human Driving: A Cut-in Prediction Study}
\author{Yu (Fred) Song$^1$}

\date{
	$^1$Department of Civil and Architectural Engineering and Construction Management, University of Wyoming \\ \texttt{ysong@uwyo.edu}\\%
}

\begin{document}
%\linenumbers
\maketitle

\begin{abstract}
Capturing the strategic decision-making inherent in competitive human driving is critical for autonomous vehicle safety and traffic simulation. This study demonstrates that game-theoretic Inverse Reinforcement Learning (IRL) provides a robust framework for this challenge. We present a comprehensive analysis comparing data-driven IRL models against an established physics-based game-theoretic approach for predicting aggressive, safety-critical cut-in lane changes. Using the high-fidelity highD dataset, we systematically develop and evaluate a series of IRL models with increasing feature complexity. Our results reveal significant advantages: the best-performing IRL models achieve an overall prediction accuracy exceeding 75\% while maintaining a Cut-In precision up to 51.0\% and recall up to 49.0\%. This represents a significant improvement over the established physics-based benchmark, which achieved only 4.4\% precision in these high-stakes scenarios. The analysis reveals a clear trade-off: incorporating granular, instantaneous features yields higher precision, while adding temporal consistency features maximizes recall. These findings suggest that IRL-based models can effectively bridge the gap between microscopic driver intent and macroscopic safety outcomes, providing a more reliable foundation for modeling interactions in mixed-autonomy environments.

\end{abstract}

\input{intro}

\input{lit_review}

\input{cut_in_game}

\input{calibration_validation}

\input{empirical}

\input{conclusion}

%	\newpage
\bibliography{refs}

\appendix
\input{appendix}

\end{document}

%% file: intro.tex
\section{Introduction}

Discretionary lane changes (DLCs) encompass a wide spectrum of behaviors, but a critical subset for traffic analysis involves aggressive, close-proximity maneuvers known as ``cut-ins'' \cite{gao_discretionary_2022, xie_modeling_2019}. Unlike a standard lane change into an ample gap, a cut-in is distinctly defined as a maneuver into a gap smaller than typically accepted safety margins, compelling an immediate and often forceful braking reaction from the trailing vehicle in the target lane \cite{fu_human-like_2019, wang_analysis_2019}. These events are characterized by their safety-critical nature, operating with minimal time-to-collision \cite{luo_risk_2023, xie_modeling_2019}; their non-cooperative and competitive dynamics, where one driver's gain directly imposes a cost on another; and their disproportionate impact on traffic stability, often acting as the catalyst for flow-disrupting shockwaves \cite{gao_discretionary_2022, lu_modeling_2022}. Because these characteristics are so distinct from routine lane changes, they cannot be accurately captured by generic models and demand a specialized approach. Therefore, understanding and modeling the high-stakes, strategic negotiation inherent in cut-ins is crucial for improving traffic safety, mitigating congestion, and developing robust decision-making for autonomous vehicles \cite{chen_human-centered_2019, remmen_cut-scenario_2018}.

Predicting these aggressive cut-in maneuvers is critical for autonomous vehicle systems to safely and efficiently navigate mixed traffic environments. Understanding when and how human drivers decide to execute such high-stakes actions is essential for developing robust autonomous driving algorithms that can anticipate and respond to competitive human behavior. Prior work has tackled this using several frameworks. These include traditional rule-based models that follow a rigid logic (e.g., \cite{kesting2007general}) and, more recently, sophisticated game-theoretic models that treat the interaction as a game of bounded rationality (e.g., \cite{arbis_game_2019, wang_modeling_2022}). While these game-theoretic models are a major advancement, their utility functions are often pre-specified based on physics-based models rather than learned from data. The unique, interactive nature of cut-ins requires a framework that not only models the strategic decision-making process but also grounds those decisions in an empirically learned utility function.

Recent advances in Inverse Reinforcement Learning (IRL) show great promise for modeling complex human decision-making by learning reward functions directly from observed behavior \cite{ng_algorithms_2000, arora_survey_2021}. Unlike traditional supervised learning, IRL seeks to understand the underlying motivations that drive human actions. This makes IRL particularly suitable for modeling the competitive dynamics of cut-ins, where drivers must rapidly weigh competing objectives like safety, efficiency, and comfort in high-pressure situations.

The highD dataset \cite{krajewski_highd_2018} provides a rich source of naturalistic driving data, which includes numerous instances of the competitive cut-in maneuvers central to this study. Containing detailed trajectory information from German highways with precise vehicle positions, velocities, and accelerations, this dataset offers a valuable opportunity to develop and validate sophisticated predictive models using realistic, high-stakes driving scenarios.

In this paper, we bridge the gap between these two advanced frameworks: game theory and IRL. As established, prior work has either employed game theory with pre-specified utility functions (e.g., \cite{wang_modeling_2022, chen_combining_2024}) or used IRL to model driving behaviors without a formal game equilibrium structure (e.g., \cite{sun_probabilistic_2018, zhou_modeling_2024}). Our work represents one of the first systematic efforts to integrate these two paradigms specifically for predicting competitive cut-in maneuvers. We present the systematic development of four IRL game-theoretic models of increasing complexity, all built on the Quantal Response Equilibrium (QRE) framework and trained on high-fidelity highD data. Our models progress from base model with core instantaneous features and a non-strategic interacting vehicle, to an enhanced model that adds granular binary features, interaction terms, and a strategic interacting vehicle, and further to an advanced model incorporating temporal consistency features (e.g., rolling means, trends) to capture the interaction's history. Finally, we test an hybrid model that fuses data-driven features with physics-based predictive simulations generated by a calibrated Intelligent Driver Model (IDM). Lacking a direct comparison model in the literature, we validate our approach against the Wang et al. (2022) model \cite{wang_modeling_2022}. We selected this as the most relevant benchmark because it models the parent class of general DLCs and, critically, shares the same underlying QRE game-theoretic framework, allowing for a rigorous comparison of the utility function derivation methods (IRL-learned vs. physics-based). We demonstrate that our specialized, data-driven IRL approach is superior for capturing the nuanced negotiations that define these critical events.

%% file: lit_review.tex
\section{Literature Review}
Modeling DLCs has been a persistent challenge in traffic flow theory. DLCs are maneuvers initiated by drivers to improve their driving conditions, such as overtaking a slower vehicle, moving to a faster lane, or avoiding a heavy vehicle. The term ``discretionary'' contrasts with ``mandatory'' lane changes (MLCs), which are necessitated by roadway geometry or traffic control. The decision process for a DLC is multi-staged, often conceptualized as a driver evaluating the possibility, necessity, and desirability of the maneuver before searching for an acceptable gap. In essence, DLCs are motivated by a driver's desire for a better state, rather than an immediate need, and their frequency and nature significantly influence overall traffic efficiency and safety.

This complex decision-making process is particularly critical for a high-stakes subset of DLCs known as cut-ins, often defined as a lane change where the accepted time gap is less than a critical safety threshold. Such maneuvers are not rare; studies using naturalistic driving data \cite{wang_analysis_2019, xie_modeling_2019, gao_discretionary_2022} have found that cut-ins can constitute a substantial portion of all lane changes and exhibit significantly higher risk profiles than typical DLCs. These events are consistently identified as a major contributor to rear-end collisions, as the forced braking reaction from the cut-off vehicle not only impacts safety and ride comfort but also requires decelerations that are significantly higher than those in normal car-following situations \cite{wang_analysis_2019, xie_modeling_2019}. At a macroscopic level, these sudden decelerations can trigger traffic oscillations and shockwaves, degrading the stability of the traffic stream and causing strong disturbances to emerging systems like vehicle platoons \cite{sultan_modeling_2002, jin_kinematic_2010, kesting_enhanced_2010, shang_cut-ins_2020, mullakkal-babu_comparative_2022}. Due to their competitive nature, cut-ins are particularly challenging to model, as they cannot be treated as isolated decisions but must be understood as a dynamic, strategic interaction between at least two self-interested agents.

Existing studies modeling both MLCs and DLCs offer a wide spectrum of frameworks. This evolution reflects an ongoing effort to better capture the complex, strategic interactions inherent in driver decision-making processes. The paradigms have progressed from early rule-based systems, such as Gipps model \cite{gipps1986model} and its enhanced versions \cite{hidas2002modelling, kesting2007general, treiber2013traffic}, to utility-based discrete choice models which were introduced to better capture driver heterogeneity \cite{ahmed1996models, ahmed1999modeling, toledo2003modeling, toledo2005lane, choudhury2006cooperative, toledo2009state}. Other approaches, including cellular automata, Markov processes, and hazard-based models, have also been explored to model the probabilistic nature of the decision \cite{chowdhury2000statistical, maerivoet2005cellular, deng2019multilane, toledo2009state, singh2011estimation, hamdar2009modeling, wang2022lcmodelav}. More recently, the focus has shifted towards advanced game-theoretic and data-driven models capable of capturing the strategic interactions \cite{ji_review_2020}.

Prior game-theoretic MLC models offer a good framework to refer to for building a game model for DLCs and further, cut-ins. MLC studies often focus on the scenario of highway ramp merging, where the game structure is typically simpler and the features considered are less complex than in discretionary scenarios. In the early MLC game framework, such as that established by Kita \cite{kita_merginggiveway_1999} and extended by Liu et al. \cite{liu_game_2007}, the interaction is modeled as a two-player (a subject vehicle and a vehicle which the subject vehicle is merging in front of), non-zero-sum, non-cooperative game solved with a Nash Equilibrium. The players are the merging vehicle and the through vehicle in the target lane, with action spaces of \{Merge, Wait\} and \{Give way, Not give way\}, respectively. The key evolution in these early models was in the payoff function, which progressed from a simple safety metric like Time-to-Collision (TTC) in Kita's work to a more complex utility function combining both safety and mobility in Liu et al.'s model. More recent studies have introduced several key innovations to this foundational framework. These include incorporating bounded rationality through the Quantal Response Equilibrium (QRE) to model imperfect human decisions \cite{arbis_game_2019}, explicitly modeling driver heterogeneity by considering different risk attitudes \cite{chen_game-theory-based_2023}, and capturing the dynamic negotiation process by modeling the interaction as a repeated game over multiple time steps \cite{kang_repeated_2020}. Other unique advancements include applying the framework to new contexts like Connected Vehicles and AV control systems \cite{ali_game_2019, huang_game-based_2024}, and using evolutionary game theory to analyze population-level strategy evolution \cite{qu_vehicle_2025}.

Prior studies on game-theoretic DLC modeling, while less numerous than those on MLCs, have established a crucial foundation for understanding strategic lane changes. One of the earlier frameworks was proposed by Talebpour et al. \cite{talebpour_modeling_2015}, who established a two-player, non-cooperative game to model DLCs, with a unique contribution in analyzing how driver utility functions change in a connected vehicle environment with V2V communication. A significant advancement was made by Wang et al. \cite{wang_modeling_2022}, who introduced two key concepts to the DLC game: the use of QRE to model the bounded rationality of drivers, and a time-dependent rationality parameter to capture how drivers may learn or become more certain as a decision unfolds. Building on the theme of bounded rationality, recent work by Chen et al. \cite{chen_combining_2024} has pushed the theoretical frontier further by replacing QRE with a Cognitive Hierarchy Model. This more advanced framework models drivers as having different levels of strategic thinking (i.e., level-k thinking) and combines this with a Deep Markov Model to capture time dependency, aiming for more human-like behavior prediction. Other researchers have applied game theory and related concepts to different aspects of the lane-change problem, such as developing AV control strategies using Stackelberg games for Right-of-Way management in mixed traffic \cite{yu_row-based_2023}, or using parallel learning frameworks that combine reinforcement and imitation learning \cite{han_modeling_2024}.

While the game-theoretic models discussed previously provide a robust structure for strategic interaction, they rely on pre-defined utility functions. We now turn to IRL, a machine learning framework designed to solve the inverse problem: inferring an agent's underlying utility function by observing their ``expert'' behavior. The application of IRL in traffic modeling, therefore, represents a paradigm shift, moving from manually specified rules to a data-driven approach that infers human intent directly from observations. Foundational work by Ng and Russell \cite{ng_algorithms_2000} established the core principle of IRL: to recover the underlying reward function that an expert agent is optimizing, given its observed actions. This makes IRL exceptionally well-suited for modeling complex human decision-making in driving, as it avoids subjective assumptions about driver priorities. A comprehensive survey by Arora and Doshi \cite{arora_survey_2021} highlights the evolution of IRL methods and their growing applicability to real-world problems like autonomous driving. In the context of driving behavior, researchers have applied IRL to learn nuanced, human-like reward functions that balance competing objectives such as safety, comfort, and efficiency. Existing work has focused on learning stochastic driver behaviors from naturalistic data, demonstrating IRL's ability to capture the probabilistic nature of human choices \cite{ozkan_inverse_2021, huang_driving_2022}. To handle the complexity and high dimensionality of driving, some recent studies have integrated IRL with deep learning. For example, \cite{zou_inverse_2018} used a neural network to approximate the reward function for car-following behavior, while \cite{fernando_deep_2021} employed a deep IRL approach for more accurate long-term behavior prediction. Other advanced methods include using adversarial IRL to better mimic human driving styles and multi-task IRL to predict fine-grained behaviors like acceleration and steering simultaneously \cite{sackmann_modeling_2022, nishi_fine-grained_2020}.

Several studies have specifically applied IRL to model interactive driving scenarios like the lane changes central to our paper. Sun et al. \cite{sun_probabilistic_2018} used a hierarchical IRL framework to make probabilistic predictions of driver intent during interactions, while more recent work by Zhou and Chen \cite{zhou_modeling_2024} applied IRL to specifically model the distinct decision logic of mandatory versus discretionary lane changes. The output of IRL is not only for prediction, but also used to develop human-like AV control systems, such as customizing an automated lane change system to match a specific driver's preferences or planning segmented lane change trajectories that consider safety and comfort \cite{liu_inverse_2022, sun_inverse_2022}. Furthermore, as AVs are expected to share roadways with human drivers, IRL is also used to model how human drivers and other road users adapt their behavior when interacting with them in mixed-autonomy environments \cite{wen_modeling_2023, huang_conditional_2023, alozi_how_2024}.

Synthesizing the prior work highlights a clear trajectory in driver modeling towards capturing strategic interaction and bounded rationality, with game theory and IRL emerging as powerful, complementary frameworks. Game theory, as employed in numerous MLC and DLC studies (e.g., \cite{kita_merginggiveway_1999, liu_game_2007, wang_modeling_2022}), provides an essential structure for modeling interactive decisions. However, these models often rely on utility functions defined by pre-specified physical models or assumptions, which may not fully capture the complex preferences underlying human choices, even when sophisticated equilibrium concepts like QRE are used. Conversely, IRL offers a robust, data-driven method to learn these nuanced utility functions directly from observations, as demonstrated in various driving behavior studies. Yet, many existing IRL applications in driving behavior modeling lack a formal game-theoretic equilibrium framework to explicitly account for the strategic interdependence and mutual anticipation between drivers in competitive scenarios like cut-ins.

Our work bridges this gap by integrating these two powerful paradigms. We leverage the structural advantages of a two-player, non-cooperative game solved with QRE to capture the strategic nature of cut-ins, but critically, we define the players' utility functions using IRL to infer them directly from observed expert driving data. This synthesis is vital for safety-critical applications, where performance cannot be measured by overall accuracy alone. For a rare but dangerous event like a cut-in, the trade-off between Recall (detecting true threats to prevent collisions) and Precision (avoiding false positives that cause unnecessary, erratic braking) is paramount. Much of the existing game-theoretic prediction literature has relied on overall accuracy as the primary performance measure (e.g., \cite{wang_modeling_2022, chen_combining_2024}), which is often misleading in highly imbalanced datasets. Our approach, in contrast, allows for an empirically grounded model of boundedly rational, strategic decision-making that is specifically tailored to—and can be evaluated on—these critical safety metrics, representing a key methodological advancement. The following sections detail our proposed models built on this integrated framework.

%% file: cut_in_game.tex
\section{The Cut-in Game}
While a cut-in is a type of DLC, its unique characteristics demand a specialized modeling approach. For this study, we define a cut-in scenario not as any DLC, but specifically as a maneuver initiated within a critically sized gap, with a speed difference, that is large enough to potentially encourage overtaking, existing between the interacting Subject Vehicle (SV) and Target Vehicle (TV). A formal definition detailing all criteria for identifying these scenarios is provided as follows:

\begin{definition}[Cut-In Scenario]
    A Cut-In Scenario is defined as an interaction between a Subject Vehicle (SV) in an original lane and a Target Vehicle (TV) in an adjacent target lane, characterized by the following conditions persisting for a continuous duration of at least 1.0 second:
    \begin{enumerate}
        \item \textbf{Strategic Gap:} An available time gap, $T_{avail}$, exists between the TV and its leader (the New Lead Vehicle, NLV) such that $1.0s < T_{avail} < 4.5s$.
        \item \textbf{Motivational Speed Difference:} The absolute speed difference between the SV and TV, $|\Delta v_{SV,TV}|$, is within the range $2 \text{ mph} (0.9 \text{ m/s}) < |\Delta v_{SV,TV}| < 20 \text{ mph} (8.9 \text{ m/s})$.
        \item \textbf{Proximity:} The SV's longitudinal position, $p_{SV}(t)$, is between the TV and the NLV, specifically within the zone defined by $p_{TV}(t) - 8 \text{ ft} (2.4 \text{ m}) < p_{SV}(t) < p_{NLV}(t)$.
    \end{enumerate}
    The outcome of the scenario is labeled as a ``Cut-In'' if the SV completes a lane change into the target lane within a 3.0-second lookahead window following the end of these conditions; otherwise, it is labeled as a ``Stay''.
    \label{def:cutin}
\end{definition}

\begin{figure}[th!]
    \centering
    \includegraphics[width=1\textwidth]{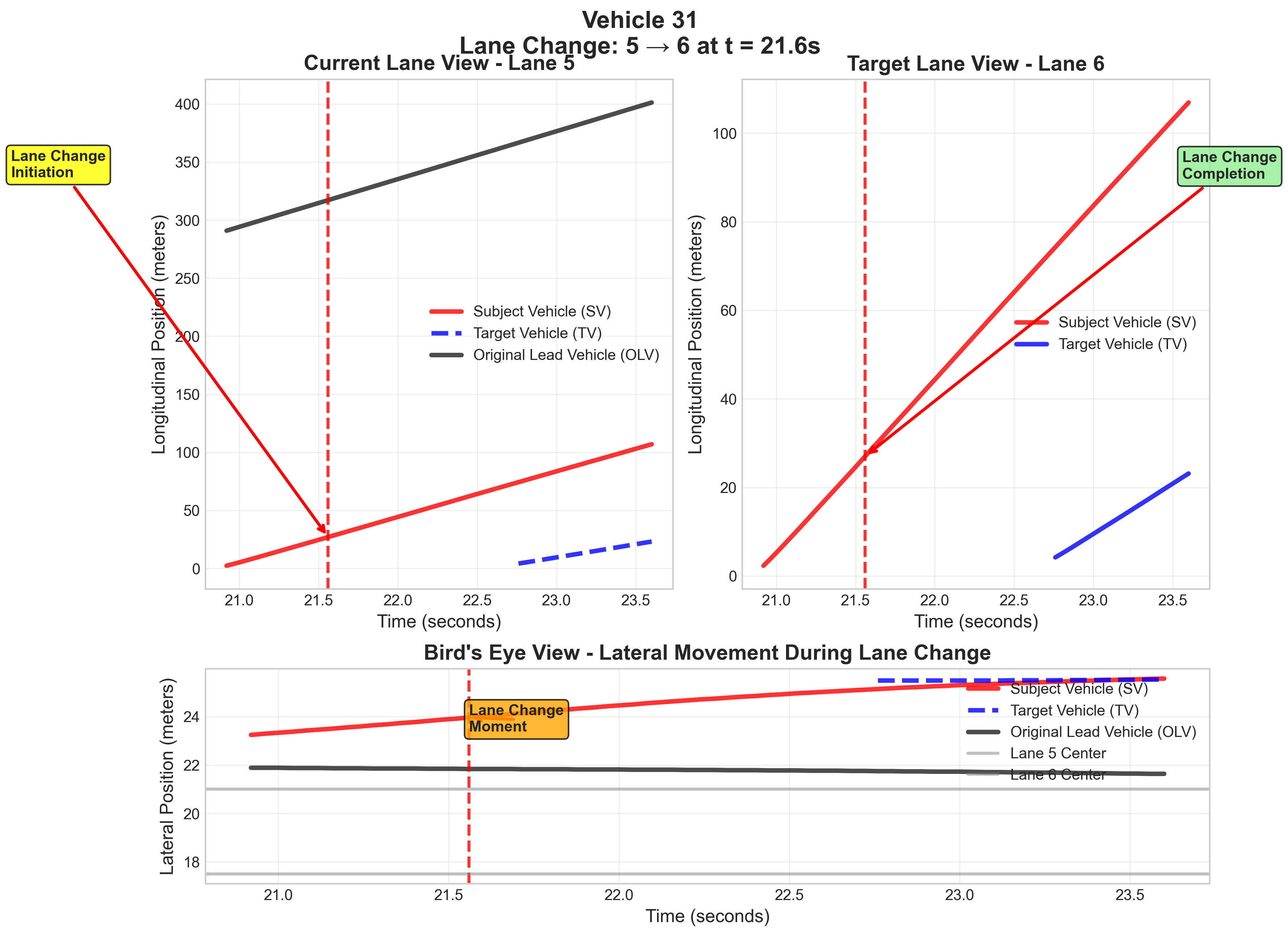}
    \caption{Example of a cut-in trajectory from the highD dataset. The Subject Vehicle (SV, red) maneuvers into a tight gap ahead of the Target Vehicle (TV, blue), illustrating the close-proximity interaction characteristic of these events.}
    \label{fig:trajectory_example}
\end{figure}

Figure \ref{fig:trajectory_example} shows an example trajectory plot from the highD dataset illustrating a cut-in maneuver. Unlike a typical DLC into a large, safe gap, this type of assertive, competitive action prioritizes the SV's benefit, forces a reaction from the TV, and necessitates a more sophisticated modeling framework and a more granular utility function than those for general DLCs. We model the cut-in as a two-player, non-cooperative game where the driver utility function is inferred from observed data via IRL and the game is solved using QRE.

\subsection{Players, State Variables, and Action Spaces}
\begin{figure}[th!]
    \centering
    \includegraphics[width=0.8\textwidth]{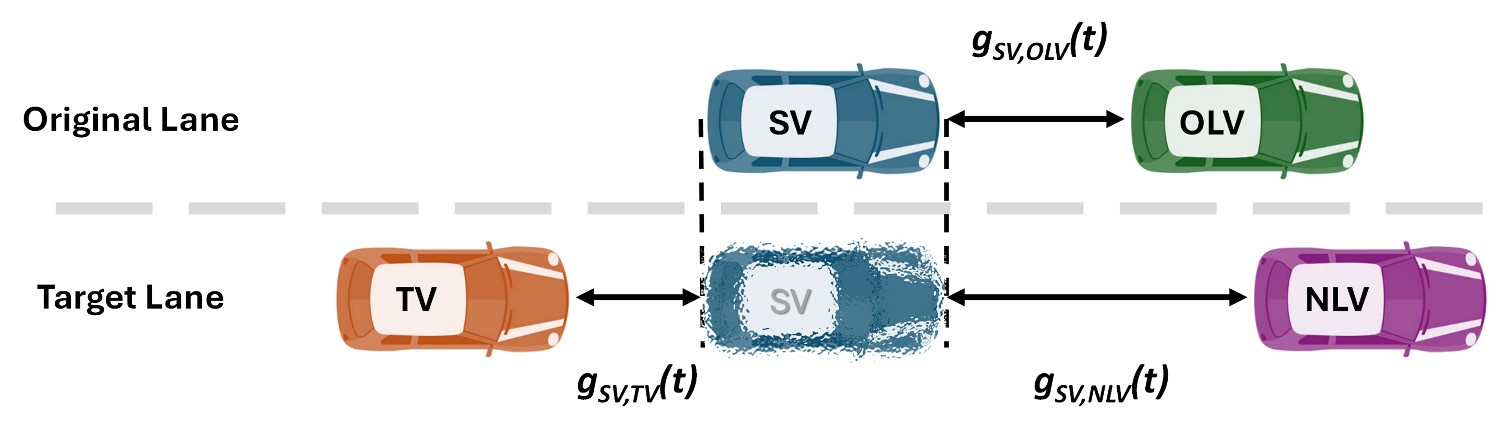}
    \caption{Schematic of the cut-in game setup, illustrating the four key vehicles: the SV, the TV being cut off, the OLV, and the NLV.}
    \label{fig:schematic}
\end{figure}

We state our model of the SV considering a lane change maneuver in the cut-in game decision process in this section. The cut-in game models the decision-making process of a SV considering a lane change maneuver, interacting with a TV. While SV and TV are the primary strategic players, the broader traffic context is defined by two other key vehicles: the SV's current leader, the Original Lead Vehicle (OLV), and its potential new leader in the target lane, the NLV, as illustrated in Figure \ref{fig:schematic}. The system state at time \(t\) includes the physical states of all relevant vehicles and contextual information about the driving environment. The physical state of any vehicle \(i\) at time \(t\) is captured by vector \(\mathbf{x}_i(t) = [p_i(t), v_i(t), a_i(t), l_i(t)]\), where the elements are the vehicle's longitudinal position, velocity, acceleration, and length, respectively.

Within this context, each player chooses an action from a discrete set. The SV's action space is \(\mathcal{A}_{SV} = \{\text{Cut-in}, \text{Stay}\}\), representing the decision to either execute the maneuver or remain in the current lane. Correspondingly, the TV selects from its action space \(\mathcal{A}_{TV} = \{\text{Yield}, \text{Maintain}\}\), representing its choice to either decelerate to accommodate the SV or maintain its current trajectory.

The system state is defined as:
\begin{equation}
\mathbf{s}(t) = [\mathbf{x}_{SV}(t), \mathbf{x}_{TV}(t), \mathbf{x}_{OLV}(t), \mathbf{x}_{NLV}(t)]
\end{equation}
where \(\mathbf{x}_{SV}(t)\) is the physical state of SV; \(\mathbf{x}_{TV}(t)\) is the physical state of TV (vehicle being cut off; \(\mathbf{x}_{OLV}(t)\) is the physical state of OLV (leader in current lane); and \(\mathbf{x}_{NLV}(t)\) is the physical state of NLV (leader in target lane)

From this state vector, all relational variables such as gaps (\(g_{i,j}\)) and relative velocities (\(\Delta v_{i,j}\)) are calculated and considered in the decision-making of SV and TV. For example:
\begin{equation}
g_{SV,OLV}(t) = p_{OLV}(t) - p_{SV}(t) - l_{SV}
\end{equation}
\begin{equation}
\Delta v_{SV,TV}(t) = v_{SV}(t) - v_{TV}(t)
\end{equation}

\subsection{Game Formulation}
The cut-in interaction is modeled as a two-player, non-cooperative game:
\[G = \langle \{SV, TV\}, \mathcal{S}, (\mathcal{A}_{SV}, \mathcal{A}_{TV}), (U_{SV}, U_{TV}) \rangle\]
where:
\begin{itemize}
    \item[] $\{SV, TV\}$ is the set of players: the Subject Vehicle and the Target Vehicle.
    \item[] $\mathcal{S}$ is the set of all possible states $\mathbf{s}(t)$, representing the traffic environment at time $t$.
    \item[] $\mathcal{A}_{SV} = \{a_{cut\text{-}in}, a_{stay}\}$ is the action set for the Subject Vehicle.
    \item[] $\mathcal{A}_{TV} = \{a_{yield}, a_{maintain}\}$ is the action set for the Target Vehicle.
    \item[] $U_{SV}$ and $U_{TV}$ are the utility functions for the SV and TV, respectively, which map a state and a pair of actions to a real-valued utility score, $U_{SV}(\mathbf{s}(t), a_{SV}, a_{TV}) \in \mathbb{R}$ and $U_{TV}(\mathbf{s}(t), a_{SV}, a_{TV}) \in \mathbb{R}$.
\end{itemize}
At each time step $t$ within an identified choice situation, the system is in a state $\mathbf{s}(t) \in \mathcal{S}$, defined by the surrounding traffic environment. The players choose actions from their respective action sets: $a_{SV} \in \mathcal{A}_{SV} = \{a_{cut\text{-}in}, a_{stay}\}$ and $a_{TV} \in \mathcal{A}_{TV} = \{a_{yield}, a_{maintain}\}$. The outcome for each player is evaluated by utility functions, $U_{SV}(\mathbf{s}(t), a_{SV}, a_{TV})$ and $U_{TV}(\mathbf{s}(t), a_{SV}, a_{TV})$, which depend on the current state and the actions chosen by both players. Operating under the assumption of bounded rationality modeled via QRE, the SV and TV aim to choose their action $a_{SV} \in \mathcal{A}_{SV}$ and $a_{TV} \in \mathcal{A}_{TV}$ that probabilistically favor maximizing their expected utilities, $E[U_{SV}(a_{SV} | \mathbf{s}(t))]$ and $E[U_{TV}(a_{TV} | \mathbf{s}(t))]$. This expected utility considers the uncertainty in the other player's action, weighted by their choice probability $P_{SV}(a_{SV} | \mathbf{s}(t))$ and $P_{TV}(a_{TV} | \mathbf{s}(t))$:

\begin{equation}
    E[U_{SV}(a_{SV} | \mathbf{s}(t))] = \sum_{a_{TV} \in \mathcal{A}_{TV}} P_{TV}(a_{TV} | \mathbf{s}(t)) \cdot U_{SV}(\mathbf{s}(t), a_{SV}, a_{TV})
\end{equation}
\begin{equation}
    E[U_{TV}(a_{TV} | \mathbf{s}(t))] = \sum_{a_{SV} \in \mathcal{A}_{SV}} P_{SV}(a_{SV} | \mathbf{s}(t)) \cdot U_{TV}(\mathbf{s}(t), a_{SV}, a_{TV})
\end{equation}

The primary objective of this game model is then to predict the strategic decision of the SV by computing the probability, $P_{SV}(a_{cut\text{-}in} | \mathbf{s}(t))$, that results from this QRE process, given the state $\mathbf{s}(t)$. Note that determining this probability for the binary action space $\mathcal{A}_{SV}$ is equivalent to learning the SV's policy $\pi: \mathcal{S} \mapsto \Delta(\mathcal{A}_{SV})$, which maps each state $\mathbf{s} \in \mathcal{S}$ to a probability distribution over the SV's actions. While calculated at each time step, this probabilistic output ultimately predicts the overall outcome of the entire choice situation event.

\subsection{Modeling Choice}
To systematically evaluate model performance, we compare our IRL game-theoretic models against the Wang et al. game-theoretic DLC model \cite{wang_modeling_2022}. We select the Wang et al. model as a benchmark because it shares the same core theoretical foundation: both approaches model the interaction as a two-player, non-cooperative game and employ QRE to account for the bounded rationality of drivers. This common framework allows for a direct comparison. However, the critical distinction lies in how the utility functions are derived and calibrated. The Wang et al. model utilizes a physics-based approach, manually specifying utility as a linear combination of simulated outputs from the IDM and the Deceleration Rate to Avoid Crash (DRAC). While they calibrated the weights of these predefined components and the parameters of their QRE model using NGSIM data, the fundamental structure of the utility function itself was assumed a priori. In contrast, our approach employs IRL to learn the utility function directly from observed driver behavior in the highD dataset.

While the QRE game-theoretic structure provides a robust framework for modeling the strategic interaction, the central challenge lies in defining the utility function that accurately reflects a driver's preferences in a high-stakes cut-in scenario. To address this, we employ IRL. Instead of manually specifying the utility function based on theoretical models like IDM and DRAC and calibrating its weights separately, IRL works backward from observed ``expert behavior'' $\mathcal{D} = \{ (\mathbf{s}_i, a_i) \}$ (i.e., the state-action pairs from cut-in decisions in the training data) to infer the underlying utility function, typically parameterized by weights $\vec{\theta}$, that the drivers were likely optimizing. The core principle is to find the parameters $\vec{\theta}$ that best explain the observed choices, often by maximizing the likelihood of the expert data given the utility function and a choice model (like QRE). Conceptually, this can be represented as finding the parameters $\vec{\theta}^*$ that maximize the probability of observing the expert actions in the corresponding states:
\begin{equation}
    \vec{\theta}^* = \underset{\vec{\theta}}{\arg\max} \ P(\mathcal{D} | U(\mathbf{s}, a; \vec{\theta}), \text{QRE}) 
\end{equation}
This data-driven approach avoids subjective assumptions about utility structure and allows the model to learn complex and often subtle trade-offs that human drivers make between competing objectives such as safety, efficiency, and comfort. By integrating IRL into the game-theoretic framework, we aim to create empirically grounded utility structures and, consequently, a more realistic and predictive driver behavior model.

Each IRL model builds upon the previous one by incorporating more complex feature sets and reasoning mechanisms, to allow for a clear analysis of how different components, from basic physical states to predictive simulations, contribute to model performance. The following is a brief overview of the IRL models' features, and more details about model formulation are introduced in the next section.

\begin{itemize}
    \item[] \textbf{IRL Base Model}: This model serves as the foundational data-driven approach. Its utility function is constructed from a focused set of 10 core features that capture the instantaneous state of the interaction. These include core metrics for feasibility, safety, comfort, and efficiency, along with simple binary flags and stay-specific metrics. This model intentionally excludes temporal features and interaction terms to establish a simple baseline.

    \item[] \textbf{IRL Enhanced Model}: This model builds upon the Base model by introducing more complex conditional logic. It augments the feature set with 6 granular binary features (e.g., for ``dangerous'' or ``risky'' safety situations) and 3 targeted interaction terms. A key enhancement is the introduction of a strategic TV, with distinct utility functions for ``Yield'' and ``Maintain'' actions.

    \item[] \textbf{IRL Advanced Model}: This model represents the most sophisticated version, whose primary innovation is the incorporation of temporal consistency features. These features include rolling means, trends, and a ``sustained opportunity'' indicator, allowing the model to learn from the recent history of the interaction. The feature set is further expanded with additional granular metrics and a total of 7 targeted interaction terms.

    \item[] \textbf{IRL Hybrid IDM Model}: This model creates a hybrid that combines data-driven features with a forward-looking, simulative component. It augments the feature set with a set of new predictive features generated from a short-term (2.0s to 4.0s) simulation using the Intelligent Driver Model (IDM) \cite{treiber2013traffic}, similar to the implementation in the Wang et al. model. A feature selection algorithm is used to identify the most statistically informative predictive features, such as the minimum predicted gap and the maximum predicted deceleration forced upon the TV.
\end{itemize}

While both our IRL models and the Wang et al. model operate within a general game-theoretic framework, they are distinguished by their unique utility function designs. The following section detail the specific formulations for the Wang et al. benchmark model and our IRL models.

\subsection{Utility Formulations}

\subsubsection{Wang et al. Benchmark Model}
This model serves as a baseline from the literature, representing a physics-based, game-theoretic approach to modeling DLCs \cite{wang_modeling_2022}. Its utility functions are derived from a simulative approach, constructed as a linear combination of metrics from established behavioral models rather than learned directly from feature interactions in data. Specifically, utilities are based on predicted acceleration from the IDM to represent efficiency, and the DRAC to represent safety.

The SV's utility, $U_{SV}(s_i, f_j)$, and the Following Vehicle's (FV) utility, $U_{FV}(s_i, f_j)$, are given by:
\begin{equation}
    U_{SV}(s_i, f_j) = \omega_1 a_{SV,ij} + \omega_2 \text{DRAC}_{SV,ij}^{LVT} + \omega_3 \text{DRAC}_{FV,ij}^{SV} + \omega_4 \text{DRAC}_{SV,ij}^{LVC} 
\end{equation}
\begin{equation}
    U_{FV}(s_i, f_j) = \mu_1 a'_{FV,ij} + \mu_2 \text{DRAC}_{FV,ij}^{LVT} + \mu_3 \text{DRAC}_{FV,ij}^{SV} 
\end{equation}
where $a$ represents the IDM predicted acceleration, DRAC terms represent safety risks, and $\omega_k, \mu_m$ ($k=1,2,3,4$, $m=1,2,3$) are utility weights. The IDM acceleration itself is a function of several parameters ($a, v_0, \delta, s_0, T, b$) describing desired speed, acceleration/deceleration limits, and desired following distance $s^*$:
\begin{equation}
a^\text{IDM} = a \left[1- \left( \frac{v}{v_0} \right)^\delta - \left(\frac{s^*(v,\Delta v)}{s} \right)^2 \right] 
\end{equation}
\begin{equation}
s^*(v, \Delta v) = s_0 + \max \left( 0, vT + \frac{v \Delta v}{2 \sqrt{ab}} \right) 
\end{equation}

Wang et al. performed empirical calibration and validation using the NGSIM dataset \cite{wang_modeling_2022}. Several sets of parameters were learned from this data: 
\begin{itemize}
    \item The IDM parameters ($\delta, v_0, s_0, T, a, b$) were calibrated from car-following trajectories.
    \item The utility weights ($\omega_k, \mu_m$) were determined using the entropy weight method applied to DLC data.
    \item The parameters of their QRE model (initial/eventual rationality $\alpha, \beta_e$, learning rate $\delta$, noise $\epsilon$) were calibrated using Maximum Likelihood Estimation on observed lane change choices.
\end{itemize}
Using this calibrated model, they reported a high prediction accuracy (over 91.5\%) on the NGSIM validation set.

However, a key limitation of the Wang et al. approach is that the fundamental structure of the utility function (i.e., the linear combination of IDM acceleration and DRAC) was assumed a priori based on theoretical considerations, rather than learned from data. While the weights and QRE parameters were calibrated empirically, this reliance on a predefined utility structure may limit the model's flexibility and ability to capture the complex, potentially non-linear trade-offs drivers make, especially in diverse datasets or for specific, nuanced maneuvers like cut-ins. This contrasts sharply with our IRL approach, which aims to infer the utility function itself from observed behavior, potentially offering greater adaptability and realism, as we will demonstrate in our performance comparison.

\subsubsection{IRL Base Model}
This foundational model serves as a robust baseline, using a set of core, instantaneous features and modeling the TV as non-strategic. The SV's utility for a ``Cut-In'' is a weighted sum of its feature vector, $\vec{\Phi}_{SV,base}$, plus a constant penalty $p_c$:
\begin{equation}
    U_{SV}(a_{cut\text{-}in}) = \vec{\alpha}_{base} \cdot \vec{\Phi}_{SV,base} + p_c
\end{equation}
where $\vec{\Phi}_{SV,base}$ is the 7-dimensional feature vector for the cut-in action, and $\vec{\alpha}_{base}$ is its corresponding learned weight vector. The constant $p_c$ is a fixed penalty for the act of cutting in.

Conversely, the utility for the ``Stay'' action is formulated as a sum of a base reward ($R_{base}$), the weighted value of its feature vector $\vec{\Phi}_{stay,base}$, and two fixed bonus terms:
\begin{equation}
    U_{SV}(a_{stay}) = R_{base} + \vec{\alpha}_{stay} \cdot \vec{\Phi}_{stay,base} + B_{safety} + B_{comfort}
\end{equation}
where $R_{base}$ is a fixed reward for staying, $\vec{\Phi}_{stay,base}$ is the 4-dimensional feature vector for the stay action (containing $S_{stay}$, $C_{stay}$, $E_{stay}$, and $\Delta\rho$), and $\vec{\alpha}_{stay}$ is its learned weight vector. $B_{safety}$ and $B_{comfort}$ are fixed, non-learned bonuses based on the core safety and comfort metrics.

The TV's utility is formulated as a linear combination of its 5-dimensional feature vector $\vec{\Psi}_{TV,base}$, governed by a weight vector $\vec{\beta}_{base}$:
\begin{equation}
    U_{TV}(a_{SV}=a_{cut\text{-}in}) = \vec{\beta}_{base} \cdot \vec{\Psi}_{TV,base}
\end{equation}
where $\vec{\Psi}_{TV,base}$ is the feature vector reflecting the state of the interaction from the TV's perspective. In this non-strategic model, the utilities for the TV's ``Yield'' and ``Maintain'' actions are identical.

\subsubsection{IRL Enhanced Model}
This model builds upon the Base model by introducing more complex conditional logic and a strategic TV. The SV's ``Cut-In'' utility function is expanded by adding a vector of 6 granular binary features ($\vec{\Phi}_{granular}$) and a vector of 3 targeted interaction terms ($\vec{\Phi}_{int,E}$) to the base feature set:
\begin{equation}
    U_{SV}(a_{cut\text{-}in}) = \vec{\alpha}_{enh} \cdot \left( \vec{\Phi}_{SV,base} \cup \vec{\Phi}_{granular} \cup \vec{\Phi}_{int,E} \right) + p_c
\end{equation}
where $\vec{\alpha}_{enh}$ is the new 17-dimensional learned weight vector corresponding to the expanded feature set, and $p_c$ is the constant penalty. The SV's ``Stay'' utility is also enhanced by adding a new learned safety bonus, $B_{learned\_safety} = \alpha_7 \cdot \mathbf{1}_{\{\text{is\_stay\_safe}\}}$.

A key change is the introduction of distinct ``Yield'' and ``Maintain'' utilities for the TV, allowing it to act strategically. Both utilities are governed by a 6-dimensional weight vector $\vec{\beta}_{E}$ and feature vector $\vec{\Psi}_{TV,E}$, but they differ in their learned weight for lane density ($\beta_5$ vs. $\beta_6$):
\begin{equation}
    U_{TV}(a_{yield}) = \vec{\beta}_{E} \cdot \vec{\Psi}_{TV,E} = \beta_1 S + \beta_2 E + \beta_3 C_{TV} + \beta_4 + \beta_5 \Delta\rho 
\end{equation}
\begin{equation}
    U_{TV}(a_{maintain}) = \vec{\beta}_{E} \cdot \vec{\Psi}_{TV,E} + (\beta_6 - \beta_5)\Delta\rho
\end{equation}
where $S$ and $E$ are core safety/efficiency metrics, $C_{TV}$ is a new comfort penalty for the TV's own acceleration, and $\beta_4$ is an intercept term.

\subsubsection{IRL Advanced Model}
This model represents the most sophisticated version, incorporating temporal consistency features. The SV's ``Cut-In'' utility function is governed by a 25-dimensional weight vector, $\vec{\alpha}_{adv}$, and a comprehensive feature vector, $\vec{\Phi}_{SV,adv}$, which combines enhanced core metrics, all granular features, 7 temporal features ($\vec{\Phi}_{temporal}$), and 7 interaction terms ($\vec{\Phi}_{int,A}$):
\begin{equation}
    U_{SV}(a_{cut\text{-}in}) = \vec{\alpha}_{adv} \cdot \vec{\Phi}_{SV,adv} + p_c
\end{equation}
The ``Stay'' utility builds upon the Enhanced model's function by adding a learned temporal bonus, $B_{temporal}$:
\begin{equation}
    U_{SV}(a_{stay}) = U_{SV,Enhanced}(a_{stay}) + B_{temporal}
\end{equation}
where $B_{temporal} = 0.2 \cdot (\alpha_{18} F_{roll} + \alpha_{19} S_{roll} + \alpha_{20} E_{roll})$. The TV's strategic utility is also enhanced with an 8-dimensional weight vector $\vec{\beta}_{adv}$ and a vector of temporal features $\vec{\Psi}_{temporal} = \{Sust_{opp}, F_{roll}\}$, allowing it to react to the stability of the SV's behavior:
\begin{equation}
    U_{TV}(a_{yield}) = U_{TV,Enhanced}(a_{yield}) + \vec{\beta}_{adv} \cdot \vec{\Psi}_{temporal} + \beta_8 \rho_{penalty}
\end{equation}
\begin{equation}
    U_{TV}(a_{Maintain}) = U_{TV,Enhanced}(a_{Maintain}) + \vec{\beta}_{adv} \cdot (1 - \vec{\Psi}_{temporal}) + \beta_8 \rho_{penalty}
\end{equation}

\subsubsection{IRL Hybrid IDM Model}
This model merges a set of 13 core data-driven features ($\vec{\Phi}_{base,H}$) with a vector of forward-looking, simulative features ($\vec{\Phi}_{pred}$) generated by the IDM. The SV's utility is a weighted sum of these two feature sets, with distinct predictive features for each action:
\begin{equation}
    U_{SV}(a_{cut\text{-}in}) = \vec{\alpha}_{base,H} \cdot \vec{\Phi}_{base,H} + \vec{\alpha}_{pred} \cdot \vec{\Phi}_{pred, cut\text{-}in}
\end{equation}
\begin{equation}
    U_{SV}(a_{stay}) = \vec{\alpha}_{base,H} \cdot \vec{\Phi}_{base,H} + \vec{\alpha}_{pred} \cdot \vec{\Phi}_{pred, stay}
\end{equation}
where $\vec{\Phi}_{pred, \cdot}$ are the vectors of predictive features (e.g., $g_{pred, cut\text{-}in}$, $a_{pred, TV, min}$) and $\vec{\alpha}_{pred}$ are their learned weights. The TV's utility function uses a 6-feature vector $\vec{\Psi}_{TV,H}$ and a unique opposing logic:
\begin{equation}
    U_{TV}(a_{yield}) = \vec{\beta} \cdot \vec{\Psi}_{TV,H}
\end{equation}
\begin{equation}
    U_{TV}(a_{maintain}) = \vec{\beta} \cdot (1 - \vec{\Psi}_{TV,H})
\end{equation}
where $\vec{\Psi}_{TV,H}$ includes core metrics and binary flags (e.g., $\mathbf{1}_{\{g_{large}\}}$, $\mathbf{1}_{\{\text{TTC}_{crit}\}}$), testing a direct trade-off in the TV's decision.

\subsection{Quantal Response Equilibrium}
The game is solved using the QRE, a solution concept from behavioral game theory that accounts for the bounded rationality of human decision-makers \cite{wang_modeling_2022}. Unlike the classical Nash Equilibrium, which assumes players are perfect optimizers, QRE models a more realistic scenario where players are prone to making errors. It posits that actions with higher expected utility are chosen more frequently, but all actions have a non-zero probability of being selected.

In our framework, the model solves a sequential game under the QRE assumption, where the SV (as the leader) anticipates the TV's (the follower's) response. This hierarchical process involves three main steps:

\begin{enumerate}
    \item \textbf{TV's Response Model:} The SV first models the TV's boundedly rational response. The probability that the TV will ``Yield'' is calculated based on the relative utilities of its available actions, governed by a rationality parameter $\lambda$:
    \begin{equation}
        P_{TV}(a_{yield} | \mathbf{s}(t)) = \frac{\exp(\lambda \cdot U_{TV}(a_{yield}, t))}{\exp(\lambda \cdot U_{TV}(a_{yield}, t)) + \exp(\lambda \cdot U_{TV}(a_{maintain}, t))}
    \end{equation}
    
    \item \textbf{SV's Expected Utility:} The SV then calculates the expected utility of cutting in by weighting the outcomes of the TV's potential responses. The theoretical expected utility $E[U_{SV}]$ for the ``Cut-In'' action is:
    \begin{equation}
    \label{eq:sv_expected_utility}
    \begin{split}
        E[U_{SV}(a_{cut\text{-}in} | \mathbf{s}(t))] = &P_{TV}(a_{yield} | \mathbf{s}(t)) \cdot U_{SV}(a_{cut\text{-}in}, a_{yield}) + \\
        &(1 - P_{TV}(a_{yield} | \mathbf{s}(t))) \cdot U_{SV}(a_{cut\text{-}in}, a_{maintain})
    \end{split}
    \end{equation}
    A key simplifying assumption in our current IRL models (Base, Enhanced, and Advanced) is that the SV's utility for the cut-in action is calculated based on the instantaneous state $\mathbf{s}(t)$ and is modeled as being independent of the TV's subsequent reaction. Therefore, we set $U_{SV}(a_{cut\text{-}in}, a_{yield}) = U_{SV}(a_{cut\text{-}in}, a_{maintain}) = U_{SV}(\mathbf{s}(t))$. This simplification means the expected utility calculation (Eq. \ref{eq:sv_expected_utility}) effectively reduces to $E[U_{SV}(a_{cut\text{-}in})] = U_{SV}(\mathbf{s}(t))$.
    
    \item \textbf{SV's Choice Probability:} Finally, the SV makes its own boundedly rational decision by comparing the expected utility of cutting in against the utility of staying:
    \begin{equation}
        P_{SV}(a_{cut\text{-}in} | \mathbf{s}(t)) = \frac{\exp(\lambda E[U_{SV}(a_{cut\text{-}in} | \mathbf{s}(t))])}{\exp(\lambda E[U_{SV}(a_{cut\text{-}in} | \mathbf{s}(t))]) + \exp(\lambda U_{SV}(a_{stay}, t))}
    \end{equation}
\end{enumerate}

The rationality parameter $\lambda$ dictates the degree of randomness in decision-making, where $\lambda \to \infty$ implies perfect rationality and $\lambda \to 0$ implies random choice. In our IRL framework, the effect of $\lambda$ is implicitly learned and absorbed into the utility function weights $\vec{\theta}$ during calibration.

%% file: calibration_validation.tex
\section{Model Calibration and Validation}

\subsection{Data Preparation}
All models were calibrated (i.e., their parameters were fit to the data) and subsequently validated using the highD dataset, a large-scale collection of naturalistic vehicle trajectories recorded on German highways. The dataset contains precise vehicle positioning and kinematic information from 60 distinct highway segments, captured by a drone at a frequency of 10Hz. We selected the highD dataset over other sources, such as the NGSIM dataset used in the Wang et al. benchmark, due to its superior data quality. While valuable, datasets like NGSIM are known to contain significant measurement noise \cite{montanino_trajectory_2015}; for example, some analyses report that raw NGSIM trajectories contain substantially higher acceleration noise and a greater frequency of unphysical jerk values compared to drone-based datasets like highD \cite{krajewski_highd_2018}. This known noise issue is why the Wang et al. study \cite{wang_modeling_2022}, which serves as our benchmark, explicitly used a reconstructed version of the NGSIM data to mitigate these problems before calibration. The high precision of the highD dataset is essential for our study, as it provides a more reliable ground truth without this intensive pre-processing, which is critical for modeling aggressive, safety-critical cut-ins that depend on accurate measurements of small changes in gaps, relative velocities, and lateral accelerations.

To prepare the data for model calibration and validation, we extracted discretionary lane change events using a systematic processing pipeline. First, we identified all potential ``choice situations'' by applying the rigorous filtering criteria set forth in Definition \ref{def:cutin}. Following this identification, we applied a Savitzky-Golay filter \cite{schafer2011savitzky}, a common signal processing technique used to smooth noisy data by fitting a low-degree polynomial to sequential windows of data, to the raw trajectories to derive stable and physically plausible kinematic profiles. From these smoothed trajectories, we then calculated the core behavioral metrics required by our models: Feasibility (F), Safety (S), Comfort (C), and Efficiency (E). The final step was labeling the outcome. As detailed in Definition \ref{def:cutin}, each choice situation was classified as a ``Cut-In'' or ``Stay'' based on the SV's actions within a 3.0-second lookahead window. The final processed dataset contains 586,031 samples (approximately 70\%) for training and 294,514 samples (approximately 30\%) for testing, with an overall class distribution of approximately 21.5\% ``Cut-In'' and 78.5\% ``Stay'' events.

\subsection{Model Calibration}
Each IRL model was calibrated using Maximum Likelihood Estimation (MLE) to find the set of utility function parameters, $\vec{\theta}$, that best explains the observed driver decisions in the training data. The objective is to maximize the log-likelihood function, which is equivalent to minimizing the negative log-likelihood, or cross-entropy loss, over all $N$ observed choice situations. The optimization was performed using the L-BFGS-B algorithm, a quasi-Newton method chosen for its efficiency in solving large-scale, parameter-rich problems. A key feature of this algorithm is its support for box constraints, which we used as a form of regularization by enforcing bounds on each parameter (e.g., $\theta_j \in [-2, 2]$) to prevent overfitting and ensure model stability.

A central challenge in this problem is the natural class imbalance between ``Stay'' events and the much rarer ``Cut-In'' events. To address this, we replaced the standard cross-entropy loss with a Focal Loss function, which focuses the training on hard-to-classify examples by down-weighting the loss contributed by well-classified samples. The Focal Loss is defined as:
\begin{equation}
    FL(p_t) = -\alpha_t (1 - p_t)^\gamma \log(p_t)
\end{equation}
where $p_t$ is the model's predicted probability for the ground-truth class, $\gamma$ is the focusing parameter, and $\alpha_t$ is a class-balancing weight. We further tuned the model's sensitivity to the minority class by applying an ``aggressiveness factor'', an additional weight applied specifically to the loss from ``Cut-In'' samples, to find the optimal balance between precision and recall for these critical events.

The final model parameters are the result of a systematic hyperparameter optimization process. This involved not only tuning the aggressiveness factor but also evaluating the impact of different feature sets, including the inclusion of targeted interaction terms and temporal features with varying rolling window sizes. Furthermore, to mitigate the effects of class imbalance at the data level, we employed propensity score matching methods to create more balanced training batches during calibration. The key hyperparameters and settings are summarized in Table \ref{tab:hyperparameters}.

\begin{table}[th!]
\centering
\caption{Key Hyperparameters and Calibration Settings}
\label{tab:hyperparameters}
\begin{tabular}{lll}
\hline
\textbf{Parameter} & \textbf{Description} & \textbf{Value / Range} \\
\hline
\multicolumn{3}{l}{\textbf{Optimization \& Regularization}} \\
Algorithm & Optimization method for minimizing loss & L-BFGS-B \\
Parameter Bounds & Box constraints on each learned weight $\theta_j$ & [-2, 2] \\
\hline
\multicolumn{3}{l}{\textbf{Loss Function}} \\
Focusing Parameter ($\gamma$) & Controls the down-weighting of easy examples & 2.0 \\
Balancing Weight ($\alpha_t$) & Balances the importance of positive/negative examples & 0.25 \\
Aggressiveness Factor & Additional weight for the minority (Cut-In) class & Tuned in [1.5, 5.0] \\
\hline
\multicolumn{3}{l}{\textbf{Model \& Data Parameters}} \\
Temporal Window ($W$) & Look-back window for rolling mean features & 5 timesteps (0.5s) \\
Lookahead Horizon & Forward-looking window to classify event outcome & 3.0 seconds \\
\hline
\end{tabular}
\end{table}

\subsection{Model Validation}
It is crucial to note that the prediction framework of our IRL models is event-based, not timestep-based. The fundamental unit of analysis and prediction is the ``choice situation'' identified during the data preparation phase. For each complete choice situation, which can span several seconds, the model generates a probabilistic output representing the likelihood of a ``Cut-In'' for that entire event. Consequently, the model predicts the driver's strategic intention for the situation, rather than the tactical action at any specific moment. This distinction is critical for interpreting the performance metrics presented in this section: metrics such as Accuracy, Precision, and Recall evaluate the model's ability to correctly classify the final outcome of the entire event, not the driver's action at each discrete timestep.

The process for generating a probabilistic prediction for each timestep within a test set choice situation is detailed in Algorithm \ref{alg:prediction_process}. The algorithm begins with the calibrated model parameters ($\vec{\alpha}$ for the SV, $\vec{\beta}$ for the TV). For every timestep $t$ in the test data, it first observes the current system state $\mathbf{s}(t)$, calculates the corresponding feature vectors ($\vec{\Phi}(t)$ for the SV, $\vec{\Psi}_{TV}(t)$ for the TV), and standardizes these features using the previously learned scaling parameters. Next, using the calibrated weights, the algorithm computes the raw utility values for each potential action available to the SV and TV at that specific moment ($U_{SV}(a_{cut\text{-}in}, t)$, $U_{SV}(a_{stay}, t)$, $U_{TV}(a_{yield}, t)$, $U_{TV}(a_{maintain}, t)$). The core of the prediction lies in solving the QRE. It estimates the TV's behavior by calculating the probability that the TV will yield, $P_{TV}(a_{Yield}, t)$, and then calculates the SV's expected utility for cutting in, $E[U_{SV}(a_{cut\text{-}in}, t)]$. Finally, the SV's probability of choosing to cut in at this timestep, $P_{SV}(a_{cut\text{-}in}, t)$, is calculated by comparing the expected utility of cutting in against the utility of staying, using the logistic quantal response function.

\begin{algorithm}[th!]
\caption{Timestep Prediction Process}
\label{alg:prediction_process}
\begin{algorithmic}[1]
\Require Calibrated model parameters $\vec{\alpha}, \vec{\beta}$; optimal decision threshold $\tau^*$
\For{each timestep $t$ in the test set}
    \State Observe the system state $\mathbf{s}(t)$
    \State Calculate the feature vectors $\vec{\Phi}(t)$ and $\vec{\Psi}_{TV}(t)$ from $\mathbf{s}(t)$ and standardize them
    
    \Statex \Comment{\textit{Calculate raw utilities for each player and action}}
    \State $U_{SV}(a_{cut\text{-}in}, t) \gets \text{Calculate using } \vec{\alpha} \text{ and } \vec{\Phi}(t)$
    \State $U_{SV}(a_{stay}, t) \gets \text{Calculate using } \vec{\alpha} \text{ and } \vec{\Phi}(t)$
    \State $U_{TV}(a_{yield}, t) \gets \text{Calculate using } \vec{\beta} \text{ and } \vec{\Psi}_{TV}(t)$
    \State $U_{TV}(a_{maintain}, t) \gets \text{Calculate using } \vec{\beta} \text{ and } \vec{\Psi}_{TV}(t)$

    \Statex \Comment{\textit{Solve the QRE to find the SV's cut-in probability}}
    \State Calculate the TV's yield probability $P_{TV}(a_{yield}, t)$:
    $$ P_{TV}(a_{yield}, t) \gets \frac{\exp(\lambda U_{TV}(a_{yield}, t))}{\exp(\lambda U_{TV}(a_{yield}, t)) + \exp(\lambda U_{TV}(a_{maintain}, t))} $$
    
    \State Calculate the SV's expected utility for cutting in $E[U_{SV}(a_{cut\text{-}in}, t)]$:
    $$ E[\cdot] \gets P_{TV}(a_{yield}, t) \cdot U_{SV}(a_{cut\text{-}in}, t) + (1 - P_{TV}(a_{yield}, t)) \cdot U_{SV}(a_{cut\text{-}in}, t) $$
    
    \State Calculate the SV's final probability of cutting in $P_{SV}(a_{cut\text{-}in}, t)$:
    $$ P_{SV}(a_{cut\text{-}in}, t) \gets \frac{\exp(\lambda E[U_{SV}(a_{cut\text{-}in}, t)])}{\exp(\lambda E[U_{SV}(a_{cut\text{-}in}, t)]) + \exp(\lambda U_{SV}(a_{stay}, t))} $$

    \Statex \Comment{\textit{Generate final binary classification using optimal threshold}}
    \If{$P_{SV}(a_{cut\text{-}in}, t) > \tau^*$}
        \State Predict $\hat{y}(t) \gets 1$ (Cut-In)
    \Else
        \State Predict $\hat{y}(t) \gets 0$ (Stay)
    \EndIf
\EndFor
\end{algorithmic}
\end{algorithm}

Due to the significant class imbalance in the dataset, a standard decision threshold of 0.5 is suboptimal for classifying events based on the output probability $P_{SV}(a_{cut\text{-}in}, t)$. Therefore, we implemented a decision threshold optimization step to identify the ideal threshold, $\tau^*$, that maximizes the F1-Score for the minority ``Cut-In'' class on the validation set. This optimal threshold is found by evaluating the F1-score across a range of potential thresholds:
\begin{equation}
    \tau^* = \underset{\tau \in [0,1]}{\arg\max} \left( \text{F1-Score}(\tau) \right)
\end{equation}
Final binary predictions, $\hat{y}(t)$, for each timestep are then generated by comparing the model's output probability against this optimal threshold, as shown in the final steps of Algorithm \ref{alg:prediction_process} ($\hat{y}(t)=1$ for a ``Cut-In'' if $P_{SV}(a_{cut\text{-}in}, t) > \tau^*$, and $\hat{y}(t)=0$ for a ``Stay'' otherwise).

Model performance was then quantified using a suite of standard classification metrics derived from a confusion matrix, which categorizes predictions against the ground-truth labels. The four components of this matrix are True Positives (TP), representing correctly identified cut-ins; True Negatives (TN), representing correctly identified stay events; False Positives (FP), where a stay was incorrectly predicted as a cut-in; and False Negatives (FN), where a cut-in was missed and predicted as a stay. From these components, we calculated the primary evaluation metrics. Accuracy measures the overall fraction of correct predictions:
\begin{equation}
    \text{Accuracy} = \frac{TP + TN}{TP + TN + FP + FN}
\end{equation}
For the critical ``Cut-In'' class, we evaluated Precision, which measures the correctness of positive predictions, and Recall, which measures the model's ability to identify all actual positive instances.
\begin{equation}
    \text{Precision} = \frac{TP}{TP + FP}
\end{equation}
\begin{equation}
    \text{Recall} = \frac{TP}{TP + FN}
\end{equation}
Finally, the F1-Score was calculated as the harmonic mean of Precision and Recall, providing a single, robust metric for evaluating performance on the imbalanced ``Cut-In'' class.
\begin{equation}
    \text{F1-Score} = 2 \cdot \frac{\text{Precision} \cdot \text{Recall}}{\text{Precision} + \text{Recall}} = \frac{2TP}{2TP + FP + FN}
\end{equation}
These metrics were calculated for both the ``Cut-In'' and ``Stay'' classes to provide a comprehensive and balanced assessment of each model's predictive power.

%% file: empirical.tex
\section{Model Performance and Discussion}
A comprehensive comparison was conducted between the physics-based Wang et al. benchmark and our series of data-driven IRL models. The analysis reveals fundamental differences in methodological approach which translate directly to significant disparities in predictive performance, particularly in identifying safety-critical cut-in maneuvers.

\subsection{Comparative Model Setup}
The five models compared in this study, while all rooted in a game-theoretic framework using QRE, represent a spectrum of modeling philosophies and complexities, as summarized in Table \ref{tab:model_setup_comparison}. The Wang et al. \cite{wang_modeling_2022} model embodies a traditional, top-down approach, deriving driver utilities from established physics-based simulations (IDM and DRAC) and incorporating a time-dependent rationality parameter to model learning. In contrast, our IRL models follow a bottom-up, data-driven paradigm. The IRL Base model establishes a foundation using only instantaneous core metrics and features a non-strategic TV model. Building upon this, the IRL Enhanced model introduces granular binary features and interaction terms, along with a strategically responsive TV model. The IRL Advanced model further incorporates temporal consistency features to capture the dynamics of the interaction over time. Finally, the IRL Hybrid IDM model explores a fusion approach, augmenting data-driven features with predictive simulations from the IDM. These architectural differences, particularly in feature engineering and the modeling of the TV's behavior, lead to variations in parameter counts (ranging from 10 to 36) and are hypothesized to directly impact predictive performance, especially for the complex cut-in scenario.

\begin{table}[th!]
\centering
\caption{Comparative Setup of All Calibrated Models}
\label{tab:model_setup_comparison}
\resizebox{\textwidth}{!}{%
\begin{tabular}{l|l|llll}
\hline
\textbf{Aspect} & \textbf{Wang et al.} & \textbf{IRL Base} & \textbf{IRL Enhanced} & \textbf{IRL Advanced} & \textbf{IRL Hybrid IDM} \\
\hline
\textbf{Core Philosophy} & Physics-Based & Data-Driven & Data-Driven & Data-Driven & Data-Driven + Physics \\
\textbf{Key Features} & IDM/DRAC Sim & Core Metrics & + Granular \& Interactions & + Temporal Features & + Predictive Features \\
\textbf{TV Model} & Strategic & Non-Strategic & Strategic & Strategic & Strategic \\
\textbf{Rationality} & Time-Dependent & Fixed & Fixed & Fixed & Fixed \\
\textbf{Parameters} & 10 & 15 & 23 & 33 & 36 \\
\hline
\end{tabular}%
}
\end{table}

\subsection{Quantitative Performance Results}
\begin{figure*}[th!]
    \centering
    \includegraphics[width=\textwidth]{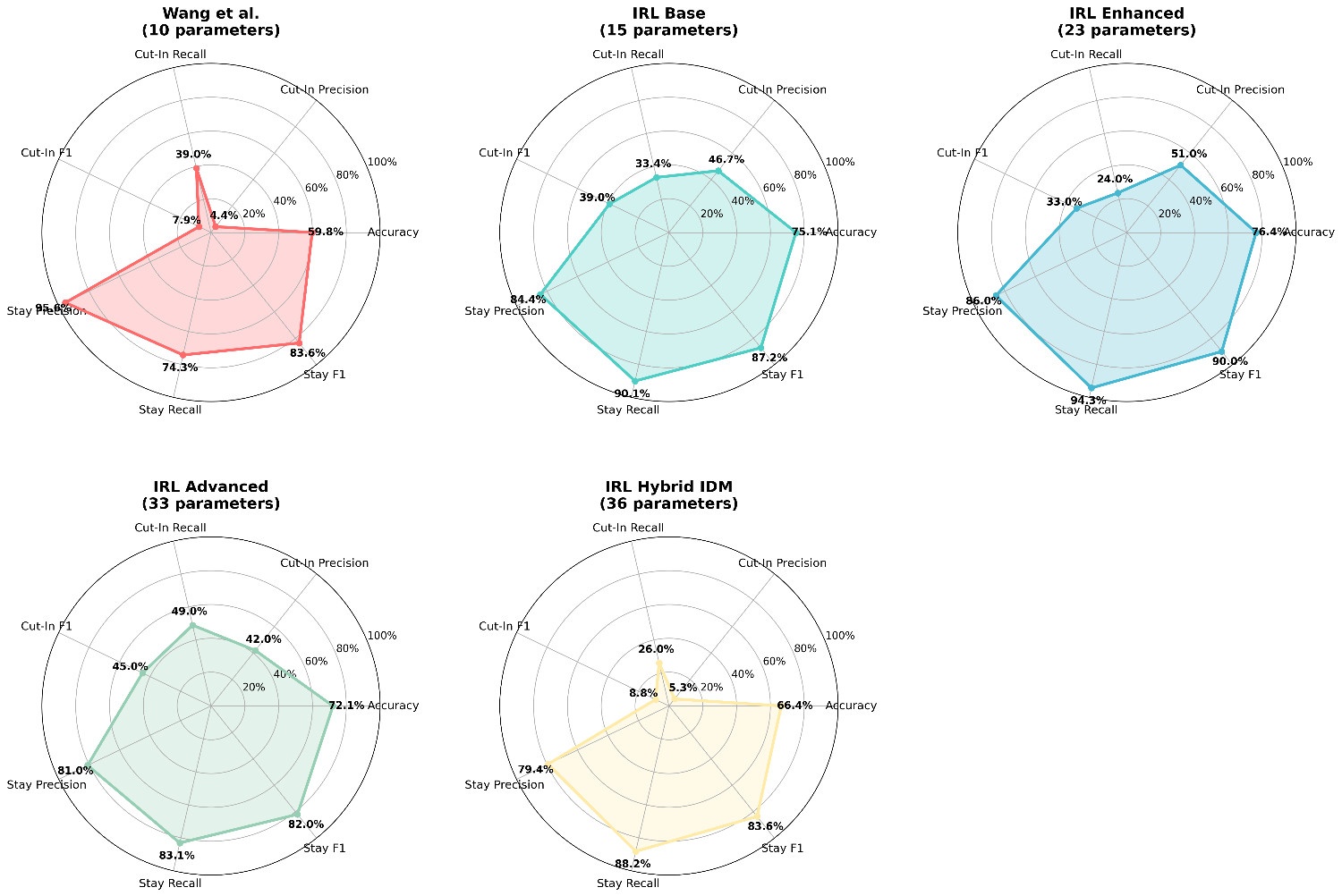}
    \caption{Performance comparison of all five models using radar charts. Each axis represents a key performance metric, with values closer to the outer edge indicating better performance. The charts visually represent the trade-offs each model makes, such as the IRL Enhanced model's strength in precision versus the IRL Advanced model's strength in recall.}
    \label{fig:radar_charts}
\end{figure*}

The predictive performance of each calibrated model was rigorously evaluated on the held-out test dataset using standard classification metrics, focusing particularly on the model's ability to handle the significant class imbalance inherent in predicting rare but critical cut-in events. The key quantitative results are presented in Table \ref{tab:model_performance_comparison}, detailing overall Accuracy alongside class-specific Precision, Recall, and F1-Scores for both ``Cut-In'' and ``Stay'' predictions. To provide a more intuitive visual comparison of the multi-dimensional performance profiles and the inherent trade-offs (e.g., between precision and recall), the results are also summarized using radar charts in Figure \ref{fig:radar_charts}. Together, the table and figure provide a comprehensive basis for assessing the strengths and weaknesses of each modeling approach, setting the stage for the detailed interpretation in the following discussion.

\begin{table}[h!]
\centering
\caption{Final Performance Metrics on the Test Dataset}
\label{tab:model_performance_comparison}
\resizebox{\textwidth}{!}{%
\begin{tabular}{l|c|c|ccc|ccc}
\hline
\textbf{Model} & \textbf{Parameters} & \textbf{Accuracy} & \textbf{Cut-In Precision} & \textbf{Cut-In Recall} & \textbf{Cut-In F1} & \textbf{Stay Precision} & \textbf{Stay Recall} & \textbf{Stay F1} \\
\hline
Wang et al. & 10 & 59.8\% & 4.4\% & 39.0\% & 0.079 & \textbf{95.6\%} & 74.3\% & 0.836 \\
\hline
IRL Base & 15 & 75.1\% & 46.7\% & 33.4\% & 0.390 & 84.4\% & 90.1\% & 0.872 \\
IRL Enhanced & 23 & \textbf{76.4\%} & \textbf{51.0\%} & 24.0\% & 0.330 & 86.0\% & 94.3\% & \textbf{0.900} \\
IRL Advanced & 33 & 72.1\% & 42.0\% & \textbf{49.0\%} & \textbf{0.450} & 81.0\% & 83.1\% & 0.820 \\
IRL Hybrid IDM & 36 & 66.4\% & 5.3\% & 26.0\% & 0.088 & 79.4\% & 88.2\% & 0.836 \\
\hline
\end{tabular}%
}
\end{table}

\subsection{Discussion of Results}
The performance evaluation reveals several key insights into the effectiveness of different modeling strategies for predicting aggressive lane changes.

\textbf{Data-driven models vastly outperform the physics-based benchmark.} The most significant finding is the clear performance gap between the data-driven IRL models and the Wang et al. model. The IRL Enhanced model achieved a Cut-In Precision of 51.0\%, over 11 times higher than the benchmark's 4.4\%. This demonstrates that while physics-based models can set theoretical safety boundaries, an empirical approach is far superior for capturing the nuanced precursors to real-world aggressive maneuvers, suggesting that human drivers' decisions are driven by a richer set of factors than purely physical constraints.

\textbf{Learned parameters provide quantitative insights into driver priorities.} The interpretability of the IRL approach allows for a data-driven view into the subconscious cost-benefit analysis of drivers. In the IRL Base model, for instance, the model learned a strong positive weight for the core safety metric ($S = 2.000$) in the cut-in utility, indicating, as expected, that perceived safety is a primary enabler of the maneuver. Simultaneously, it learned a large negative weight for efficiency ($E = -1.733$), reflecting a strong aversion to cutting in front of a faster-moving vehicle. For the ``Stay'' action, large positive weights for comfort and efficiency ($C_{stay} = 2.000$, $E_{stay} = 2.000$) quantify the significant inertia drivers exhibit when their current lane offers a smooth and unhindered ride. These learned parameters translate complex interactions into interpretable driver preferences.

\textbf{The critical nature of cut-ins places different demands on Precision and Recall based on the application.} While overall accuracy is high, this metric is misleading due to the severe class imbalance. The true challenge lies in predicting the rare, high-consequence ``Cut-In'' event. For a safety-critical application like an autonomous vehicle's (AV) warning system, Recall is paramount. A missed cut-in (a False Negative) could be catastrophic, so the system must identify the highest possible percentage of true threats, even if it generates some false alarms. This favors a high-recall model like our IRL Advanced model (49.0\% recall). Conversely, for an AV's automated planning module, Precision is vital. A high-precision model, like our IRL Enhanced model (51.0\% precision), minimizes false positives. This prevents the AV from braking unnecessarily, which would lead to an uncomfortable, inefficient, and potentially unsafe (from a following vehicle's perspective) ride. For use cases in microscopic simulation and traffic flow analysis, a model must be well-balanced to generate realistic, human-like behavior. A model with the highest F1-Score (the harmonic mean of Precision and Recall), such as our IRL Advanced model (0.450 F1-score), provides the most realistic balance between missing events and over-predicting them, making it the most suitable choice for simulating emergent traffic phenomena.

\begin{figure*}[th!]
    \centering
    % Row 1
    \subcaptionbox{Available Time Gap ($T_{avail}$)\label{fig:dist_t_avail}}{%
        \includegraphics[width=0.48\textwidth]{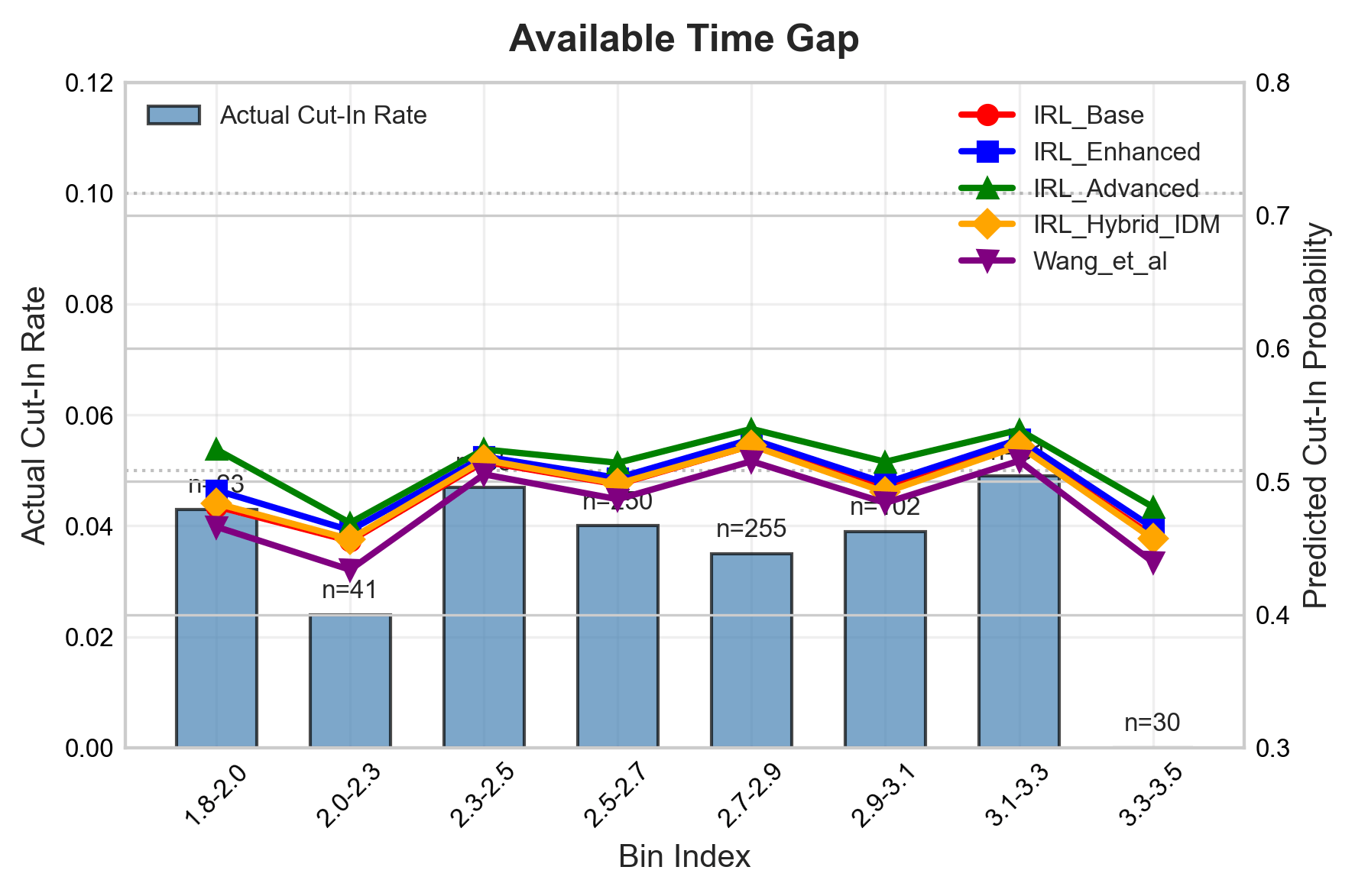}%
    }
    \hfill % Adds horizontal space between the two columns
    \subcaptionbox{Relative Velocity ($\Delta v_{TV,SV}$)\label{fig:dist_delta_v}}{%
        \includegraphics[width=0.48\textwidth]{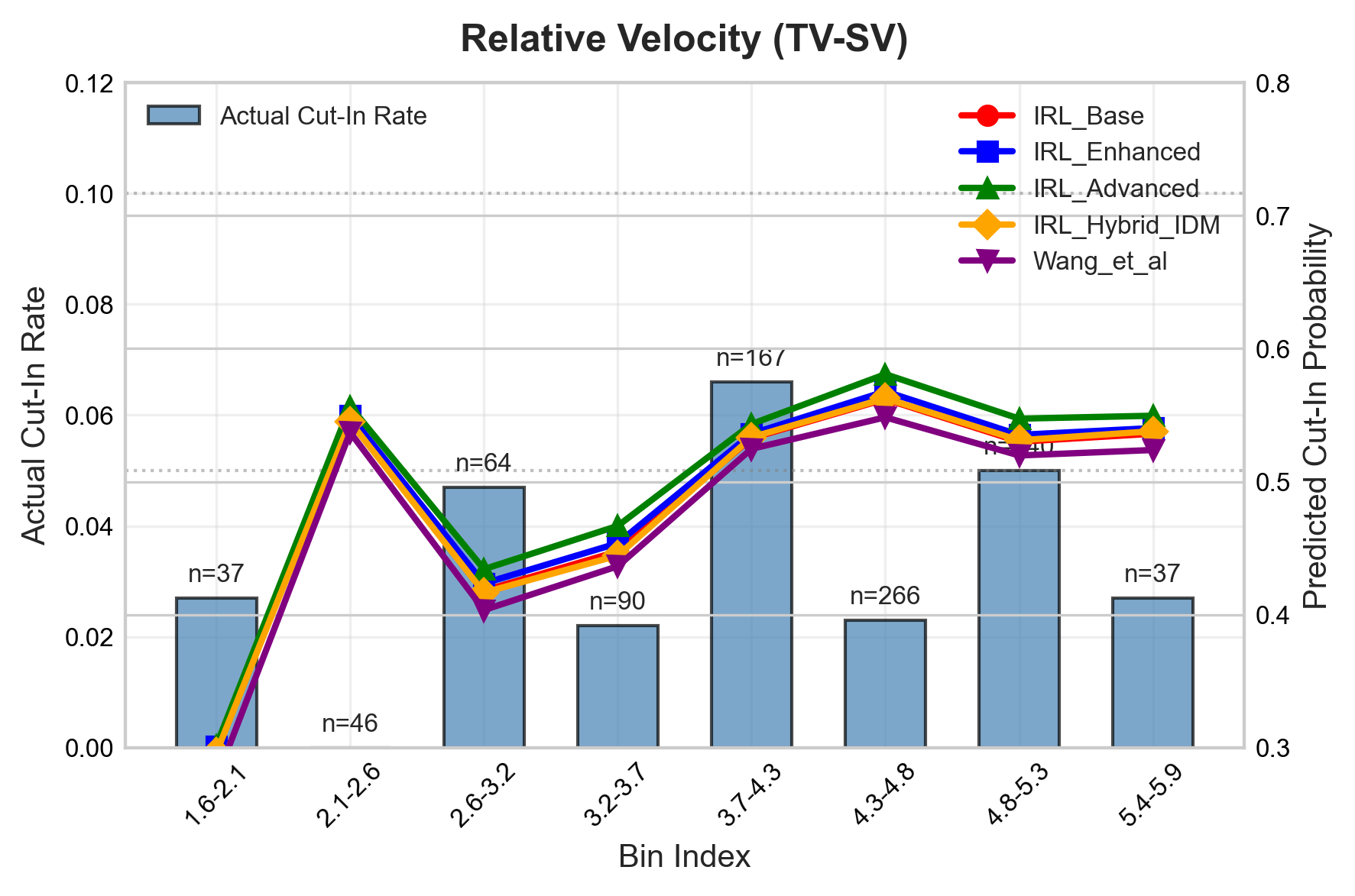}%
    }
    
    % Row 2
    \subcaptionbox{Time-to-Collision ($\text{TTC}_{SV,TV}$)\label{fig:dist_ttc}}{%
        \includegraphics[width=0.48\textwidth]{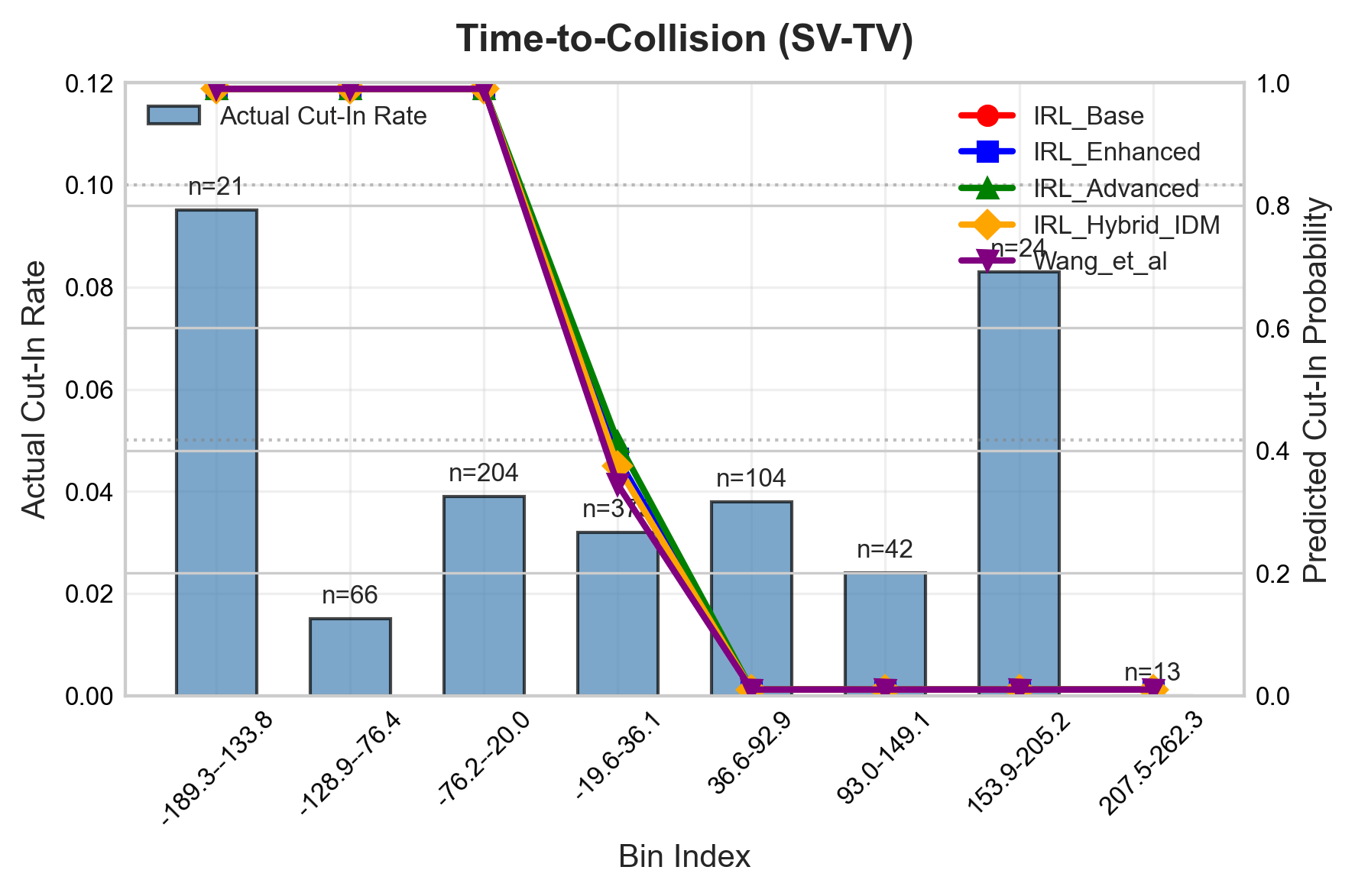}%
    }
    \hfill % Adds horizontal space between the two columns
    \subcaptionbox{Time Headway to Leader ($T_{SV,OLV}$)\label{fig:dist_t_olv}}{%
        \includegraphics[width=0.48\textwidth]{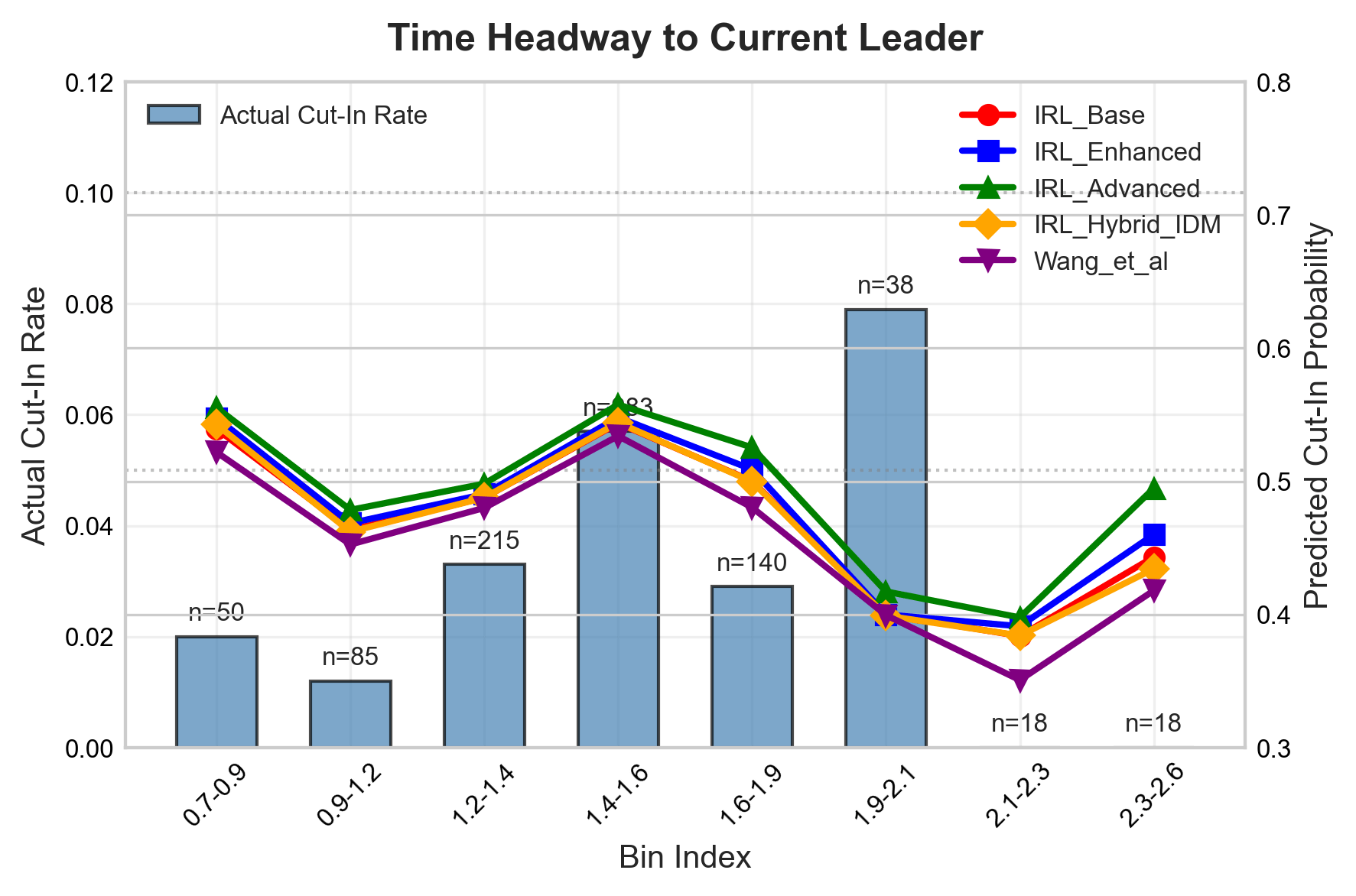}%
    }
    
    \caption{Comparison of the IRL Advanced model's predicted cut-in probability versus the actual cut-in frequency across key state variables on the test set. The blue bars represent the actual frequency of cut-ins observed in the data within each bin (left y-axis), while the orange line shows the model's average predicted probability of a cut-in for events in that bin (right y-axis).}
    \label{fig:calibration_plots}
\end{figure*}

\textbf{The model's performance reflects the inherent challenges of predicting rare, noisy human decisions.} It is important to contextualize the performance metrics. The overall accuracy of our best models (e.g., 76.4\% for IRL Enhanced) may seem modest, especially when the baseline ``always stay'' accuracy is 78.5\%. However, this is not a failure, but a direct consequence of optimizing for the rare ``Cut-In'' class. To achieve any recall for cut-ins, the model must accept some False Positives, which necessarily reduces its overall accuracy below the trivial baseline. More telling are the diagnostic plots in Figure \ref{fig:calibration_plots}, which often show the model's average predicted probability (colored lines) as higher than the actual frequency of cut-ins (blue bars). This is a common and expected outcome of training on imbalanced data. To learn to identify the rare ``Cut-In'' signal, our calibration process uses a FL function and an aggressiveness factor. These methods intentionally force the model to become highly sensitive to any features associated with the minority class. The model learns that the cost of a False Negative is high, so it assigns a higher-than-actual probability to any ambiguous situation that might become a cut-in. This leads to calibrated, but intentionally ``pessimistic'' or ``cautious'' probabilities, optimized for F1-score and Recall rather than pure probabilistic accuracy.

Furthermore, this task is characterized by a low signal-to-noise ratio. The ``signal'' is the set of observable kinematic features ($F, S, C, E$, etc.), but the ``noise'' is the high-dimensional, unobserved driver state, including factors like distraction, fatigue, mood, or individual risk tolerance. In the dataset, two scenarios with nearly identical observable features can result in different outcomes (one ``Cut-In'', one ``Stay'') due to this unobserved human element. This inherent ambiguity, or noise, places a hard upper limit on the achievable performance of any model. The 70-80\% accuracy and the probabilistic discrepancies seen in the calibration plots are a direct reflection of a model successfully finding the optimal, boundedly-rational policy in a noisy, real-world system.

\textbf{Diagnostic plots confirm the model is capturing the true signal within the noise.} Beyond aggregate metrics, it is crucial to verify that the model's predictions align with physical reality. Figure \ref{fig:calibration_plots} provides this diagnostic validation. The plots show that the model has successfully learned the underlying relationships: the predicted probability of a cut-in (colored lines) correctly trends with the actual observation (blue bars), increasing as the available time gap ($T_{avail}$) and the time-to-collision ($\text{TTC}_{SV,TV}$) decrease, and as the relative speed advantage ($\Delta v_{TV,SV}$) increases. As explained in the previous paragraph, the model's predicted probabilities are often calibrated to be higher than the actual frequencies due to the loss function's focus on the rare ``Cut-In'' class. The critical insight from these plots is not a perfect match in absolute value, but rather that the shape and direction of the predicted probabilities correctly mirror the real-world risk dynamics, confirming that the model is identifying the true, interpretable drivers of the decision.

\textbf{Incremental feature complexity provides targeted performance gains, linking feature type to prediction outcome.} The systematic model progression demonstrates how specific types of features directly influence distinct performance aspects. Adding granular binary features and interaction terms in the IRL Enhanced model yielded the largest gain in precision (from 46.7\% to 51.0\%), suggesting these features are crucial for accurately confirming the conditions necessary for a cut-in decision. Subsequently, incorporating temporal consistency features in the IRL Advanced model produced the largest gain in recall (from 24.0\% to 49.0\%), indicating that history-based features are essential for identifying the developing intent or opportunity for a cut-in, even if the instantaneous conditions are ambiguous. This highlights the complementary roles of instantaneous state representation and temporal dynamics in modeling complex driving decisions.

\textbf{Model limitations reveal the pitfalls of purely theoretical and potentially overly complex approaches.} The two worst-performing models provide valuable lessons. The Wang et al. model, despite its theoretical grounding, proved overly conservative. Its extremely high Stay Precision (95.6\%) suggests its rigid, physics-based rules effectively identify definitively safe situations but fail to capture the nuanced, negotiated risk-taking inherent in human cut-in behavior. At the other end of the complexity spectrum, the IRL Hybrid IDM model, despite having the most parameters (36), also performed poorly (5.3\% Cut-In Precision). This indicates that simply adding more complex, simulated features does not guarantee better performance; the predictive IDM features were likely too noisy, insufficiently informative for the specific task of cut-in prediction, or difficult to integrate effectively, ultimately degrading predictive power compared to the purely data-driven Advanced model.

\textbf{Several limitations warrant acknowledgment and suggest directions for future work.} The models were developed using the highD dataset, representing German highway driving; performance may vary in different traffic cultures or roadway types. Furthermore, the models rely solely on observable kinematic data and do not account for unobserved driver states like distraction, fatigue, or mood, which can influence decision-making. Lastly, our IRL models assume a fixed rationality parameter within the QRE, unlike the time-dependent parameter in the Wang et al. model; exploring dynamic rationality within the IRL framework could be a fruitful avenue for future research.

%% file: conclusion.tex
\section{Conclusions}
This study presented a comprehensive development and evaluation of IRL models for predicting aggressive cut-in lane changes, demonstrating their effectiveness compared to established, physics-based game-theoretic approaches. Through a systematic model development process, we showed how progressive feature engineering leads to significant, targeted performance gains. Our best-performing IRL models achieved superior prediction capabilities, with the Enhanced model reaching a Cut-In Precision of 51.0\% and the Advanced model achieving a Cut-In Recall of 49.0\%, both representing substantial improvements over the 4.4\% precision and 39.0\% recall of the Wang et al. benchmark.

The key contribution of this paper is the demonstration that a data-driven, game-theoretic IRL framework, grounded in rich feature engineering, is significantly more effective at capturing the nuanced decision-making behind high-stakes cut-in maneuvers than traditional physics-based models. We identified that granular, instantaneous features are crucial for achieving high precision, while temporal consistency features are essential for maximizing recall. Crucially, our IRL models provide a superior balanced performance; they not only dramatically improve the detection of critical cut-ins but also achieve a more robust prediction of ``Stay'' events, with Stay F1-scores as high as 0.900 compared to the benchmark's 0.836. This validates the IRL framework as a practically applicable solution for autonomous vehicle systems that must anticipate and safely navigate competitive human driving behavior.

The results confirm that data-driven IRL offers a robust methodology for modeling the complex, strategic interactions inherent in driving. Future work should focus on optimizing these models for real-time implementation, integrating their predictions into AV control systems to enable more proactive and human-like responses, and validating their performance across diverse traffic scenarios and driving cultures to ensure broader applicability.

%% file: appendix.tex
\section*{Appendix}

\section{Comprehensive Feature Definitions for IRL Models}
\label{app:features}

The features used in the utility functions undergo a two-step normalization process:
\begin{enumerate}
    \item \textbf{Bounding:} During feature engineering, metrics are bounded to a meaningful range using a \texttt{clip} function, e.g., $\text{clip}(x, a, b) = \max(a, \min(x, b))$.
    \item \textbf{Standardization:} All clipped features are then standardized using Standard Scaling ($z = (x_{clipped} - \mu) / \sigma$) to have a mean of 0 and a standard deviation of 1, ensuring stable model convergence.
\end{enumerate}

Table \ref{tab:all_features} details all features used across the four IRL models. Note that the Hybrid IDM model uses a distinct set of base features, which are defined differently from the other three models.

\begin{table*}[h!]
\centering
\caption{Comprehensive Feature List for all IRL Models. (B=Base, E=Enhanced, A=Advanced, H=Hybrid)}
\label{tab:all_features}
\resizebox{\textwidth}{!}{%
\begin{tabular}{llccccL{8cm}}
\hline
\textbf{Symbol} & \textbf{Feature Definition} & \textbf{B} & \textbf{E} & \textbf{A} & \textbf{H} & \textbf{Mathematical Representation / Derivation} \\
\hline
\multicolumn{7}{l}{\textbf{Category 1: SV Cut-In Core Metrics}} \\
$F$ & Feasibility of target gap & \checkmark & \checkmark & & \checkmark & $F = \text{clip}(T_{avail}/3.0, 0, 1)$ \\
$S$ & Safety for new lag vehicle (TV) & \checkmark & \checkmark & & \checkmark & $S = \text{clip}(1/\max(\text{TTC}_{SV,TV}, 0.1), 0, 10)$ \\
$C$ & Comfort penalty (longitudinal) & \checkmark & \checkmark & & \checkmark & $C = - \text{clip}(a_{SV,lon}^2, 0, 100)/100$ \\
$E$ & Efficiency gain vs TV & \checkmark & \checkmark & & \checkmark & $E = \text{clip}((v_{TV} - v_{SV})/30.0, -1, 1)$ \\
$C_{adv}$ & Comfort penalty (lon. + lat.) & & & \checkmark & & $C_{adv} = - \text{clip}((a_{SV,lon}^2 + a_{SV,lat}^2)/100, 0, 2)$ \\
$E_{adv}$ & Efficiency gain (enhanced) & & & \checkmark & & $E_{adv} = \text{clip}((v_{TV} - v_{SV})/30.0 + 0.1(v_{SV}/v_{TV}), -1, 1)$ \\
\hline
\multicolumn{7}{l}{\textbf{Category 2: SV Cut-In Binary \& Granular Features}} \\
$S_{critical}$ & Critical safety indicator & \checkmark & \checkmark & \checkmark & & $\mathbf{1}_{\{\text{TTC}_{SV,TV} < 3.0\}}$ \\
$F_{large}$ & Large gap indicator & \checkmark & \checkmark & \checkmark & & $\mathbf{1}_{\{F > 0.5\}}$ \\
$E_{huge}$ & Huge speed advantage & & \checkmark & \checkmark & & $\mathbf{1}_{\{v_{TV} - v_{SV} > 15.0\}}$ \\
$E_{mod}$ & Moderate speed advantage & & \checkmark & \checkmark & & $\mathbf{1}_{\{5.0 < v_{TV} - v_{SV} \le 15.0\}}$ \\
$S_{danger}$ & Dangerous safety indicator & & \checkmark & \checkmark & & $\mathbf{1}_{\{\text{TTC}_{SV,TV} < 2.0\}}$ \\
$S_{risky}$ & Risky safety indicator & & \checkmark & \checkmark & & $\mathbf{1}_{\{2.0 \le \text{TTC}_{SV,TV} < 3.5\}}$ \\
$F_{tiny}$ & Tiny gap indicator & & \checkmark & \checkmark & & $\mathbf{1}_{\{T_{available} < 1.5\}}$ \\
$F_{medium}$ & Medium gap indicator & & \checkmark & \checkmark & & $\mathbf{1}_{\{1.5 \le T_{available} < 3.0\}}$ \\
$E_{slight}$ & Slight speed advantage & & & \checkmark & & $\mathbf{1}_{\{0 < v_{TV} - v_{SV} \le 5.0\}}$ \\
$S_{mod}$ & Moderate safety indicator & & & \checkmark & & $\mathbf{1}_{\{3.5 \le \text{TTC}_{SV,TV} < 5.0\}}$ \\
$F_{large\_det}$ & Detailed large gap indicator & & & \checkmark & & $\mathbf{1}_{\{T_{available} \ge 3.0\}}$ \\
\hline
\multicolumn{7}{l}{\textbf{Category 3: SV Cut-In Temporal \& Interaction Features}} \\
$Sust_{opp}$ & Sustained Opportunity & & & \checkmark & & $\mathbf{1}_{\{ (F_{roll} > 0.6) \land (E_{roll} > 0.3) \land (S_{roll} < 0.5) \}}$ \\
$I_1$ & $F_{large} \times E_{huge}$ & & \checkmark & \checkmark & & Interaction Term \\
$I_2$ & $S_{danger} \times F_{tiny}$ & & \checkmark & \checkmark & & Interaction Term \\
$I_3$ & $E_{huge} \times \Delta\rho$ & & \checkmark & \checkmark & & Interaction Term \\
$I_4$ & $F_{large} \times E_{mod}$ & & & \checkmark & & Interaction Term \\
$I_5$ & $S_{risky} \times F_{medium}$ & & & \checkmark & & Interaction Term \\
$I_6$ & $Sust_{opp} \times E_{huge}$ & & & \checkmark & & Interaction Term \\
$I_7$ & $F_{trend} \times E_{trend}$ & & & \checkmark & & Interaction Term \\
\hline
\multicolumn{7}{l}{\textbf{Category 4: SV Stay Utility Features}} \\
$S_{stay}$ & Safety of staying (vs OLV) & \checkmark & \checkmark & \checkmark & \checkmark & $S_{stay} = \text{clip}(1/\max(T_{SV, OLV}, 0.1), 0, 10)$ \\
$C_{stay}$ & Comfort of staying (current $a_{SV,lon}$) & \checkmark & \checkmark & \checkmark & \checkmark & $C_{stay} = - \text{clip}(a_{SV,lon}^2, 0, 100)/100$ \\
$E_{stay}$ & Efficiency of staying (vs OLV) & \checkmark & \checkmark & \checkmark & \checkmark & $E_{stay} = \text{clip}((v_{OLV} - v_{SV})/30.0, -1, 1)$ \\
$\Delta\rho$ & Lane density change & \checkmark & \checkmark & \checkmark & & $\Delta\rho = \rho_{target} - \rho_{current}$ \\
$B_{safety}$ & Fixed safety bonus & \checkmark & \checkmark & \checkmark & & $B_{safety} = 0.5 \cdot \text{clip}(S, 0, 5)$ \\
$B_{comfort}$ & Fixed comfort bonus & \checkmark & \checkmark & \checkmark & & $B_{comfort} = 0.3 \cdot \text{clip}(-C, 0, 3)$ \\
$Stay_{safe}$ & Learned safety bonus indicator & & \checkmark & \checkmark & \checkmark & $\mathbf{1}_{\{(T_{SV,OLV} > 2.0) \land (v_{SV} - v_{OLV} > -5.0)\}}$ \\
$B_{temporal}$ & Learned temporal bonus & & & \checkmark & & $B_{temp} = 0.2 \cdot (\alpha_{18} F_{roll} + \alpha_{19} S_{roll} + \alpha_{20} E_{roll})$ \\
\hline
\multicolumn{7}{l}{\textbf{Category 5: TV Utility Features}} \\
$S, E, \Delta\rho$ & Core metrics & \checkmark & \checkmark & \checkmark & & See definitions in Category 1 and 4. \\
$C_{TV}$ & TV comfort penalty & \checkmark & \checkmark & \checkmark & & $C_{TV} = -a_{TV,lon}^2$ \\
$1$ & Intercept (Base Cooperation) & \checkmark & \checkmark & \checkmark & & $1$ \\
$F_{roll}, Sust_{opp}$ & Temporal features for TV & & & \checkmark & & See definitions in Category 3. \\
\hline
\multicolumn{7}{l}{\textbf{Category 6: Hybrid IDM Model Features (Distinct Set)}} \\
$\vec{\Phi}_{base,H}$ & 14 core data-driven features & & & & \checkmark & $F, S, C, E, \mathbf{1}_{\{g_{large}\}}, \mathbf{1}_{\{\text{TTC}_{crit}\}}, \mathbf{1}_{\{v_{adv}\}},$ \\ & & & & & & $I_{F \times E}, I_{S \times C}, I_{g \times v}, S_{stay}, C_{stay}, E_{stay}, Stay_{safe}$ \\
$\vec{\Phi}_{pred}$ & 7 types of predictive features (from IDM sim) & & & & \checkmark & $g_{pred, cut\text{-}in}, a_{pred, TV, min}, g_{pred, stay},$ \\ & & & & & & $ DRAC_{pred, TV, max}, DRAC_{pred, SV, max}, DRAC_{pred, TV, avg},$ \\ & & & & & & $DRAC_{pred, SV, avg}$ \\
$\vec{\Psi}_{TV,H}$ & 6 TV features for hybrid logic & & & & \checkmark & $S, E, \mathbf{1}_{\{g_{large}\}}, \mathbf{1}_{\{\text{TTC}_{crit}\}}, \Delta v_{TV,SV}, g_{avail}$ \\
\hline
\end{tabular}%
}
\end{table*}

The features in Table \ref{tab:all_features} are derived from several key kinematic variables calculated at each timestep $t$ from the smoothed trajectory data. The primary variables are:

\textbf{Available Time Gap ($T_{avail}$):} The time gap in the target lane between the NLV and TV. It is calculated from the longitudinal positions ($p$), vehicle lengths ($l$), and the TV's velocity ($v_{TV}$).
\begin{equation}
    g_{avail} = p_{NLV}(t) - p_{TV}(t) - l_{NLV}
\end{equation}
\begin{equation}
    T_{avail} = \frac{g_{avail}}{\max(v_{TV}(t), 0.1)}
\end{equation}
    
\textbf{Time-to-Collision ($\text{TTC}_{SV,TV}$):} The time it would take for the SV to collide with the TV if both maintained their current velocity and the SV was in the target lane. It is calculated from the longitudinal gap between them ($g_{SV,TV}$) and their relative velocity ($\Delta v_{SV,TV}$).
\begin{equation}
    g_{SV,TV} = p_{TV}(t) - p_{SV}(t) - l_{SV}
\end{equation}
\begin{equation}
    \Delta v_{SV,TV} = v_{SV}(t) - v_{TV}(t)
\end{equation}
\begin{equation}
    \text{TTC}_{SV,TV} = 
    \begin{cases} 
          \frac{g_{SV,TV}}{\Delta v_{SV,TV}} & \text{if } \Delta v_{SV,TV} > 0 \text{ and } g_{SV,TV} > 0 \\
          10.0 & \text{otherwise (no collision risk)}
    \end{cases}
\end{equation}
    
\textbf{Current Lane Time Headway ($T_{SV,OLV}$):} The time headway between the SV and its current leader (i.e., the OLV).
\begin{equation}
    T_{SV,OLV} = \frac{p_{OLV}(t) - p_{SV}(t) - l_{OLV}}{\max(v_{SV}(t), 0.1)}
\end{equation}
    
\textbf{Temporal Features ($X_{roll}, X_{trend}$):} The rolling mean features ($F_{roll}, S_{roll}, E_{roll}$) are calculated as the mean of the respective feature over a 5-timestep window (0.5s). The trend features ($F_{trend}, S_{trend}, E_{trend}$) are calculated as the first-order difference, $X_{trend}(t) = X(t) - X(t-1)$.
    
\textbf{Hybrid Model Features ($g_{pred}, a_{pred}, DRAC_{pred}$):} These features are derived from running an IDM simulation for multiple horizons (e.g., 2.0s, 3.0s, 4.0s) from the current state $t$. For example, $g_{pred, cut\text{-}in}$ is the minimum simulated gap between the SV and TV over the simulation horizon, and $a_{pred, TV, min}$ is the minimum simulated acceleration (maximum deceleration) experienced by the TV during that same simulation.

\section{Calibrated Model Parameters}

This section details the final learned weights ($\vec{\theta}$) and fixed parameters for the IRL Base Model, as derived from the calibration process described in the main text.

\begin{table}[th!]
\centering
\caption{Calibrated Learned Parameters for the IRL Base Model}
\label{tab:calibrated_params_base}
\begin{tabular}{llr}
\hline
\textbf{Utility Function} & \textbf{Parameter / Feature} & \textbf{Learned Weight ($\theta$)} \\
\hline
\multicolumn{3}{l}{\textbf{SV Cut-In Utility ($U_{SV}(a_{cut\text{-}in})$)}} \\
& Feasibility ($F$) & 1.537 \\
& Safety ($S$) & 2.000 \\
& Comfort ($C$) & -0.175 \\
& Efficiency ($E$) & -1.733 \\
& Critical Safety ($S_{critical}$) & 0.600 \\
& Large Gap ($F_{large}$) & -0.861 \\
& Lane Density Change ($\Delta\rho$) & 0.567 \\
\hline
\multicolumn{3}{l}{\textbf{SV Stay Utility ($U_{SV}(a_{stay})$)}} \\
& Safety of Staying ($S_{stay}$) & -0.586 \\
& Comfort of Staying ($C_{stay}$) & 2.000 \\
& Efficiency of Staying ($E_{stay}$) & 2.000 \\
& Lane Density Change ($\Delta\rho$) & 0.567 \\
\hline
\multicolumn{3}{l}{\textbf{TV Utility ($U_{TV}$)}} \\
& Core Safety ($S$) & -0.344 \\
& Core Efficiency ($E$) & -0.026 \\
& TV Comfort Penalty ($C_{TV}$) & 0.443 \\
& Base Cooperation (Intercept) & 0.053 \\
& Lane Density Change ($\Delta\rho$) & -0.414 \\
\hline
\end{tabular}
\end{table}

\begin{table}[th!]
\centering
\caption{Fixed Parameters in the IRL Base Model}
\label{tab:fixed_params_base}
\begin{tabular}{llr}
\hline
\textbf{Parameter} & \textbf{Description} & \textbf{Value} \\
\hline
$p_c$ & Constant penalty for executing a cut-in & -1.0 \\
$R_{base}$ & Base reward for the ``Stay'' action & 5.0 \\
$B_{safety}$ Multiplier & Fixed multiplier for the safety bonus in the Stay utility & 0.5 \\
$B_{comfort}$ Multiplier & Fixed multiplier for the comfort bonus in the Stay utility & 0.3 \\
\hline
\end{tabular}
\end{table}